\documentclass[twocolumn]{aastex631}
\usepackage{xcolor}

\newcommand{\teff}{\ensuremath{T_{\rm eff}}}
\newcommand{\logg}{\ensuremath{\log g}}

\newcommand{\logLlam}{\ensuremath{\log L_\lambda}}
\newcommand{\vsini}{\ensuremath{v\sin i}}
\newcommand{\msun}{\ensuremath{M_\odot}}
\newcommand{\AV}{\ensuremath{A_{\rm V}}}
\newcommand{\RV}{\ensuremath{R_{\rm V}}}
\newcommand{\tauage}{\ensuremath{\tau_{\rm age}}}
\newcommand{\Minit}{\ensuremath{M_{\rm init}}}

\newcommand{\kms}{\ensuremath{\,{\rm km\,s^{-1}}}}
\newcommand{\kK}{\,{\rm kK}}
\newcommand{\Nfiducial}{64{,}613}
\newcommand{\Nhot}{\Nfiducial}
\newcommand{\Nfit}{287{,}435}
\newcommand{\Nhotapprox}{65{,}000}
\newcommand{\NfidFull}{119{,}121}

\graphicspath{{figures/}}

\begin{document}

\title{Stellar parameters for $\sim\Nhotapprox$ OB Stars in the Galactic Disk from SDSS-V}

\author[0000-0003-4996-9069]{Hans-Walter Rix}
\affiliation{Max-Planck-Institut f\"ur Astronomie, K\"onigstuhl 17, D-69117 Heidelberg, Germany}

\author[0000-0003-3769-8812]{Eleonora Zari}
\affiliation{INAF--Osservatorio Astrofisico di Arcetri, Largo Enrico Fermi 5, I-50125 Firenze, Italy}

\author[0000-0001-9590-3170]{Johanna M\"uller-Horn}
\affiliation{Max-Planck-Institut f\"ur Astronomie, K\"onigstuhl 17, D-69117 Heidelberg, Germany}

\author[0000-0002-7984-1675]{Jaime Villase\~nor}
\affiliation{Max-Planck-Institut f\"ur Astronomie, K\"onigstuhl 17, D-69117 Heidelberg, Germany}

\author[0000-0002-5365-1267]{Marina Kounkel}
\affiliation{Department of Physics, University of North Florida, 1 UNF Drive, Jacksonville, FL 32224, USA}

\author[0000-0003-1822-7126]{Conny Aerts}
\affiliation{Institute of Astronomy, KU Leuven, Celestijnenlaan 200D, B-3001 Leuven, Belgium}
\affiliation{Max-Planck-Institut f\"ur Astronomie, K\"onigstuhl 17, D-69117 Heidelberg, Germany}

\author[0000-0001-8341-3940]{Mojgan Aghakhanloo}
\affiliation{Department of Astronomy, University of Virginia, Charlottesville, VA 22904, USA}

\author[0000-0003-0174-0564]{Andrew R. Casey}
\affiliation{Center for Computational Astrophysics, Flatiron Institute, 162 Fifth Avenue, New York, NY 10010, USA}

\author[0000-0003-3526-5052]{Jos\'e G. Fern\'andez-Trincado}
\affiliation{Centro de Investigaci\'on en Astronom\'ia, Facultad de Ingenier\'ia, Ciencia y Tecnolog\'ia, Universidad Bernardo O'Higgins, Av. Viel 1497, Santiago, 8370993, Chile}

\author{Lynne A. Hillenbrand}
\affiliation{Department of Astronomy, California Institute of Technology, Pasadena, CA 91125, USA}

\author[0000-0002-7955-7359]{Dhanesh Krishnarao}
\affiliation{Department of Physics, Colorado College, 14 East Cache la Poudre St., Colorado Springs, CO 80903, USA}

\author[0000-0002-3651-5482]{Jiadong Li}
\affiliation{Max-Planck-Institut f\"ur Astronomie, K\"onigstuhl 17, D-69117 Heidelberg, Germany}

\author[0000-0002-6972-6411]{Jos\'e Eduardo M\'endez Delgado}
\affiliation{Instituto de Astronom\'ia, Universidad Nacional Aut\'onoma de M\'exico, Ciudad de M\'exico, Mexico}

\author[0000-0002-7064-099X]{Dante Minniti}
\affiliation{Departamento de Ciencias F\'isicas, Universidad Andr\'es Bello, Av. Rep\'ublica 220, Santiago, Chile}

\author[0000-0002-6770-2627]{Sean Morrison}
\affiliation{Department of Astronomy, University of Illinois at Urbana-Champaign, Urbana, IL 61801, USA}

\author[0000-0001-8600-4798]{Carlos G. Rom\'an-Z\'u\~niga}
\affiliation{Instituto de Astronomía, Universidad Nacional Aut\'onoma de M\'exico, AP 106, Ensenada 22800, BC, M\'exico}

\author[0000-0002-0635-2264]{Abhinna Sundar Samantaray}
\affiliation{Astronomisches Rechen-Institut, Zentrum f\"ur Astronomie der Universit\"at Heidelberg, M\"onchhofstr.\ 12--14, D-69120 Heidelberg, Germany}

\author[0000-0002-2269-9348]{Sergio S\'anchez-Sanju\'an}
\affiliation{Instituto de Astronom\'ia, Universidad Nacional Aut\'onoma de M\'exico, Ciudad de M\'exico, Mexico}

\author[0000-0002-2090-9751]{Andreas A.\ C.\ Sander}
\affiliation{Astronomisches Rechen-Institut, Zentrum f\"ur Astronomie der Universit\"at Heidelberg, M\"onchhofstr.\ 12--14, D-69120 Heidelberg, Germany}

\author{Donald P. Schneider}
\affiliation{Department of Astronomy and Astrophysics, The Pennsylvania State University, University Park, PA 16802, USA}

\author[0000-0003-0842-2374]{Andrew Tkachenko}
\affiliation{Institute of Astronomy, KU Leuven, Celestijnenlaan 200D, B-3001 Leuven, Belgium}

\begin{abstract}
Hot, luminous stars in the Milky Way are rare, short-lived, but critical for Galactic evolution: they trace recent star
formation, dominate the radiative and mechanical feedback into the interstellar
medium, and are probes of the arguably least understood regime of stellar evolution. 
We determine physical parameters for the hot stars observed spectroscopically by SDSS-V. For each star we derive $\teff$, $\logg$,
extinction ($A_V$, $R_V$), the effective projected rotation $\vsini$, and a spectrophotometric distance by jointly
fitting their SDSS-V BOSS spectra and \emph{Gaia} XP spectrophotometry, along with their parallax.
Masses, ages, and luminosities then follow from single-star isochrones. 
We derive and provide these parameters for  $\Nfiducial$ stars from SDSS DR20 with $\teff>10\kK$ that pass our quality cuts.
We validate the stellar labels against the IACOB O-star and LAMOST surveys, as well as the SDSS-V's BOSSNet pipeline, and characterize the
sample's selection function. This sample is $\sim 100\times$ larger than existing high-resolution
spectroscopic samples of Galactic OB stars, while offering key stellar parameters, such as $\teff$,  of comparable fidelity.
\end{abstract}

\keywords{Massive stars --- Stellar atmospheres --- Surveys --- Milky Way stellar content}

\section{Introduction}\label{sec:intro}
Hot, massive stars are engines of galaxy evolution. Despite their rarity, they dominate the
ultraviolet radiation, the mechanical energy output, and the chemical enrichment that a stellar population
returns to its surroundings \citep[e.g.,][]{Naze2006, EldridgeStanway2022}. Their short lifespans ($< t_{orbit}$ in the Galaxy) make them faithful tracers of where and when star
formation has recently occurred \citep{Zari2021}. They are also the least well-understood phase of
stellar evolution: binarity \citep{Sana2012,MoeDiStefano2017}, rotation, and mass loss all leave their imprint on the
upper main sequence in ways that current models reproduce only imperfectly \citep{MaederMeynet2000,Langer2012,Smith2014}. Progress on all of these
fronts is limited in part by the lack  of \emph{large, uniformly analysed} samples of hot
stars with reliable physical parameters.

Assembling such a sample is difficult for three reasons. First, hot stars are rare and short-lived, so they
are heavily outnumbered in any magnitude-limited survey; second, they are born in dusty regions and
lie preferentially in the obscured Galactic plane, so reddening both dims them and
corrupts photometric temperature estimates. Third, Galactic OB stars have a vast range of apparent magnitudes, with the nearby and unobscured sample members being often much brighter than the ``saturation limit" of surveys designed to look mainly at faint objects. For all these reasons, hot stars have therefore remained somewhat at the margins of the large spectroscopic surveys: RAVE, APOGEE, and GALAH \citep{Steinmetz2020,Majewski2017,DeSilva2015} were designed around the cool, chemically informative stars that dominate Galactic archaeology, and their primary abundance pipelines are built on atmosphere model grids that stop well short of the OB regime \citep[e.g.][]{GarciaPerez2016}.

Where hot stars have been analysed within such surveys, it has been through separate efforts: APOGEE's OB stars, observed initially as telluric standards, have been classified semi-empirically from their $H$-band Brackett and He\,{\sc ii} lines \citep{RomanLopes2018,RamirezPreciado2020}, and assigned $\teff$ and $\logg$ across the full temperature range by a neural network trained on literature labels \citep{Sprague2022}. APOGEE's own synthetic-grid pipeline reaches only to $20\kK$ \citep{Abdurrouf2022}, so the O stars and early B stars lie beyond it; and the $H$-band spectra of hot stars are in any case sparse in diagnostics, carrying the Brackett series and, for the O stars, two He\,{\sc ii} lines. The Gaia-ESO Survey ran a dedicated analysis chain for its $\teff>7\kK$ targets \citep{Blomme2022}, but on a sample of order $10^3$ stars. Only LAMOST has produced model-based hot-star labels at scale, through a purpose-built analysis \citep{Xiang2022}; it is important to note that LAMOST's Northern survey footprint precludes observations of the inner Galaxy, where most massive stars are expected to live. Dedicated high-resolution spectroscopic surveys such as IACOB \citep{SimonDiaz2014,Holgado2018} instead deliver high-fidelity parameters of Galactic O- and early-B stars, which we take as the benchmark to compare against in Sect.~\ref{sec:val_external}, but only for samples of hundreds.

Two developments now enable a much more comprehensive spectroscopic OB star survey of the Galactic disk. 
\emph{Gaia} \citep{GaiaDR3} provides
a parallax and a low-resolution spectrophotometric (XP) spectrum for most stars with $G<17.6$, whose absolute flux calibration then ties the apparent brightness to a luminosity; 
and SDSS-V \citep{Kollmeier2026}, through its Milky Way
Mapper program, has made it a focus to target hundreds of thousands of OB-star candidates, selected
from \emph{Gaia} and 2MASS photometry by \citet{Zari2021},to observe them with its BOSS optical
spectrographs \citep[$R\simeq2000$, $3600$--$10\,400$\,\AA ;][]{Smee2013}. Neither ingredient alone is ideal to get precise, robust and accurate stellar parameters: XP
constrains the spectral energy distribution but not the surface gravity or the ionised-helium
diagnostics at $\teff >25$kK, while the BOSS spectra carry the lines but not an absolute flux scale. Fitting them
\emph{jointly}, together with the parallax, breaks many of the degeneracies that defeat either in isolation
and is the approach we take here.

This paper presents the resulting catalog. We set out to characterise
\emph{all} of the hot stars that SDSS-V has observed through MJD~61116 (2026 March~17), by deriving for each a consistent set of
atmospheric (\teff , \logg , \vsini), extinction (\AV , \RV ), and evolutionary parameters (ages \tauage\ and initial masses \Minit) via stellar isochrones, along with the most likely distance $d$. To constrain these parameters, we fit for each object the BOSS spectra, the Gaia XP spectra and the parallaxes simultaneously. From this we
define a clean hot-star sample from the initial target sample.
We treat this analysis as the data basis for the other aspects of the SDSS-V OB-star program. Therefore we present the initial sample, our method to determine stellar parameters, the resulting parameters and their validation. We defer the astrophysical exploitation, e.g., a map of recent star-formation, multiplicity or runaway stars, the clustering of recent star formation, and the large-scale kinematics of the young disc \citep{Zari2025}, to companion papers. Where we show distributions of physical quantities, we do so foremost to convey the scope and quality of the catalog along with potential limitations, not to
interpret them.

The paper is organised as follows. Sect.~\ref{sec:data} describes the data and the target sample;
Sect.~\ref{sec:fitting} the models, the joint likelihood, and the parameters we derive, including an
explicit account of what the spectral data constrain versus what parameters rely on isochrones; Sect.~\ref{sec:results}
then defines a hot-star sample with some purity tiers and validates the labels
against external and internal benchmarks; Sect.~\ref{sec:OBstar_population} presents the distribution of stellar labels across the population of the sample; Section  \ref{sec:selfunc}
characterizes the completeness of our fiducial sample \emph{vis-a-vis} the initial SDSS-V target carton \citep{Zari2021}; finally,  Sect.~\ref{sec:summary} previews some of the astrophysics that the catalog enables and summarizes.

\section{Sample and Data}\label{sec:data}
\begin{figure*}
\includegraphics[width=\textwidth]{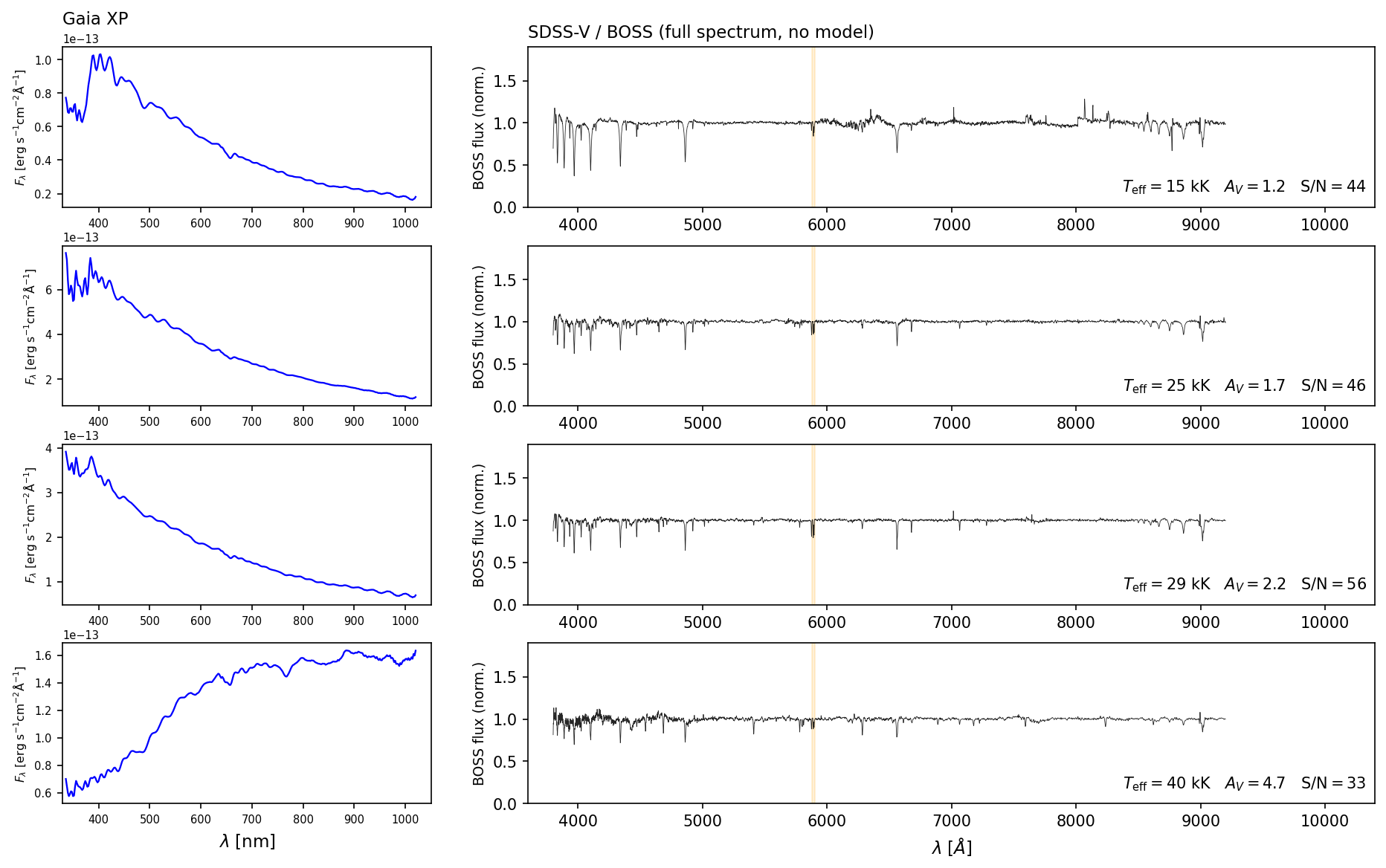}
\caption{\label{fig:data}The two spectroscopic ingredients for four example hot stars spanning
$\teff\simeq15$--$40\kK$ (data only, no model): the \emph{Gaia} XP spectral energy distribution
(left, each scaled to unit peak flux for display) and the continuum-normalised SDSS-V/BOSS diagnostic windows (right;
the interstellar Na\,D region masked from the fit is shaded). The hydrogen Balmer series and the
neutral and ionised helium lines that drive the temperature and gravity diagnostics are readily
visible and evolve systematically with $\teff$. The XP energy distribution reflects both temperature
and interstellar reddening -- the bottom star ($A_V=4.7$) is intrinsically the hottest yet appears
red because of extinction, which our joint fit disentangles.}
\end{figure*}

For each star, the raw ingredients of our analysis are an SDSS-V/BOSS optical spectrum, a \emph{Gaia}
XP spectrophotometric spectrum, and a \emph{Gaia} parallax. Fig.~\ref{fig:data} shows examples of
the two spectroscopic ingredients for hot stars before any modeling, to convey the data quality and
the diagnostic lines the fit will use.

\subsection{Target selection}\label{sec:targets}
The stars considered here are a spectroscopic subset of the parent sample of luminous hot-star
\emph{candidates} assembled by \citet{Zari2021} from \emph{Gaia} astrometry and photometry together
with 2MASS: a catalog designed to err on the side of \emph{completeness} rather than \emph{purity}, so that it
can serve as a target list for spectroscopic follow-up. SDSS-V  targets these candidates as one of its science cartons through its Milky Way Mapper program \citep{Kollmeier2026}. Fibers are assigned
operationally by the robotic focal-plane system on the Sloan 2.5\,m telescopes \citep{Gunn2006,Bowen1973} at Apache Point and Las
Campanas, subject to the survey's field and cadence planning; a given candidate is observed if and
when a fiber is available in a scheduled field covering it. Which of the parent-sample stars ends up
with a spectrum is therefore set mostly by such operational assignments rather than by a simple magnitude cut fully controlled subsampling  \citep{Blanton2026-robostrategy}. As the initial sample of \citet{Zari2021} contained many (and important) stars much brighter than SDSS-V's standard bright limit of $G \sim 13$, brighter sample members were observed with the BOSS fibers intentionally offset from the actual stars by an amount that would deliver a fiber magnitude of $G \sim 13$ for median seeing conditions \citep[see Section 4 of][]{Medan2025}. We will return to the selection function in  Sect.~\ref{sec:selfunc}. Spectra enter our analysis subject only to minimal quality assurance, namely a successful reduction by the SDSS-V BOSS pipeline (version \texttt{v6\_2\_1}). We analyze all spectra obtained through MJD~61116 (2026 March~17), keeping the input deliberately inclusive and flagging anomalous or mis-calibrated spectra only in our output catalog (Sect.~\ref{sec:products}) rather than removed.

\subsection{BOSS spectroscopy}\label{sec:boss}
SDSS-V obtains optical spectra with the BOSS spectrographs \citep{Smee2013}, covering
$\sim3600$--$10400$\,\AA\ at resolution $R\sim2000$ (see the right column of Fig.~\ref{fig:data}). Most targets are visited more than once (a median of three epochs and up to 38, though about a third are observed only once). We use both the individual-epoch reductions and a per-star \emph{coadd} of all of a star's visits. These coadds are not the BOSS pipeline's \citep{Morrison2026} standard per-field, per-night combinations; here, all epochs for any star have been shifted in a pre-processing step into the star's rest frame using the pipeline's heliocentric velocity estimate (the \texttt{XCSAO\_RV}); only then are the epochs combined into a single, higher-signal-to-noise spectrum. Rest-framing before stacking avoids additional broadening of the photospheric lines, but the coadd carries of course no useful velocity information: the velocity we fit to it is merely a ``nuisance parameter', a possible small residual scattered about zero (Sect.~\ref{sec:products}). Second, any feature that is in a  different velocity frame frame (e.g. interstellar absorption) is displaced to the \emph{negative} of the star's heliocentric velocity (Sect.~\ref{sec:byproducts}).

\begin{figure*}
\includegraphics[width=\textwidth]{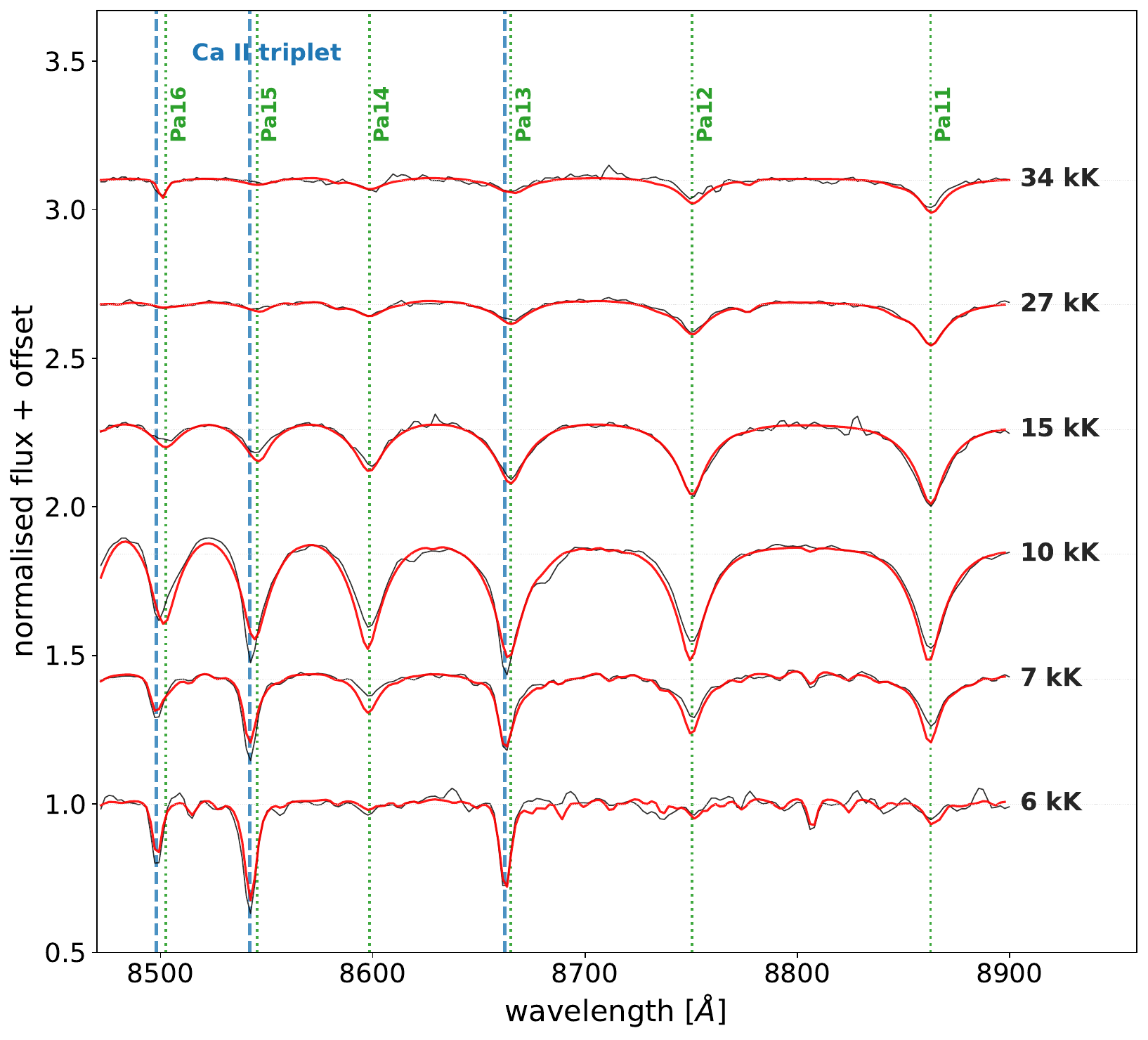}
\caption{\label{fig:paschen}The far-red Paschen $+$ Ca\,{\sc ii}-triplet window (8470--8900\,\AA) across a temperature sequence (data black, best-fit model red, offset by $\teff$). This region is rarely exploited for hot stars, yet it is both diagnostic and robust to reddening. The hydrogen Paschen lines (Pa11--Pa16) are deepest near A0 ($\sim10$--$15\kK$) and weaken monotonically upward as hydrogen ionises; the Ca\,{\sc ii} triplet contributes only at the cool end. As for the Balmer lines, the Paschen line strengths also depend on $\logg$ through pressure broadening, so this window carries gravity as well as temperature information. Because the window sits in the far red, it retains signal for heavily obscured stars that the blue diagnostics lose to extinction.}
\end{figure*}

For some stars, the BOSS spectra still contain continuum `artifacts' (wiggles or modest jumps). Therefore, we use for the spectral fitting not the full spectrum, but thirteen narrow,
continuum-normalised windows centred on temperature- and gravity-sensitive lines: twelve in the blue, spanning the Balmer and helium diagnostics, and one
in the far red (8470--8900\,\AA) covering the hydrogen Paschen series and the Ca\,{\sc ii} infrared
triplet. The H$\alpha$ and H$\beta$ regions were not assigned such a window, because they are frequently contaminated by winds and
nebular emission, often in the stars of potentially greatest interest. The red window is included because it has more information content for severely reddened objects (A$_V\gtrsim 4-5$): for the most reddened stars it carries
several times the blue signal, and its Paschen lines constrain the parameters where the blue is
photon-starved. The Paschen lines are of course temperature diagnostic in their own right
(Fig.~\ref{fig:paschen}): they are strongest near A0 and weaken steadily as hydrogen ionises with increasing \teff\ towards O stars. The interstellar Na\,D doublet and the $\lambda$8620 diffuse interstellar band, which
fall in the He\,{\sc i}\,5876 and red windows, are masked from the fit and measured separately
(Sect.~\ref{sec:byproducts}).

\subsection{Gaia XP spectrophotometry and parallaxes}\label{sec:gaia}
For every star we retrieve the \emph{Gaia}~DR3 low-resolution XP spectrum
\citep{GaiaDR3,DeAngeli2023} and the parallax. The XP continuous coefficients are sampled onto a
common wavelength grid and placed on an absolute flux scale (erg\,s$^{-1}$\,cm$^{-2}$\,\AA$^{-1}$). The primary value of these spectra lies in their \emph{spectrophotometry}, which can be linked to the
bolometric flux, or bolometric luminosity via the parallax or distance. Parallaxes are corrected for the published zero-point \citep{Lindegren2021}. The XP data provide the
primary lever on $\teff$ and the extinction through the shape of the spectral energy distribution, and on the luminosity and distance, via the spectrophotometry and parallax as observables. The large majority of our
targets have \emph{Gaia} astrometry and XP spectra. Only $\approx8\%$ of the BOSS-observed OB
candidates lack usable XP spectral information, which makes a joint analysis possible for
the sample as a whole. For the hot, intrinsically luminous stars of interest this removes mainly
distant, heavily reddened objects; we fold the limit into the selection-function analysis of
Sect.~\ref{sec:selfunc}.

\section{Models and Fitting}\label{sec:fitting}

\begin{figure*}
\centering
\includegraphics[width=0.85\textwidth]{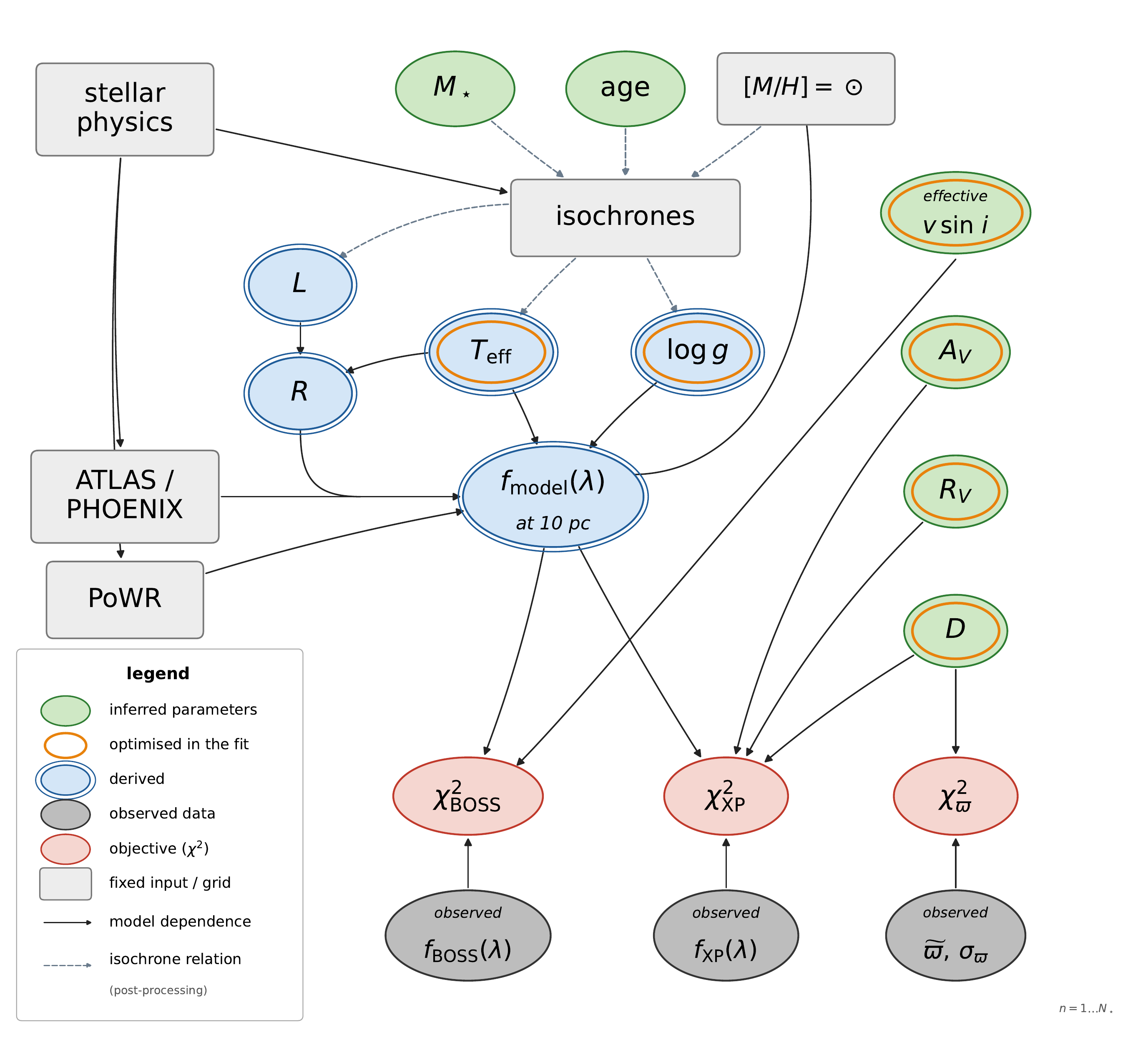}
\caption{\label{fig:schematic}Directed graphical model describing our fitting approach that combines the information of BOSS's detailed spectral features, Gaia's XP spectrophotometry, and the luminosity constraint that comes from the stars' parallaxes, which then requires simultaneous fitting of the dust extinction. Green ellipses are the 
inferred parameters, that ultimately describe the observed properties of the system; orange inner rings mark the six quantities that the optimiser actually varies:
$\teff$, $\logg$, the extinction pair $A_V$ and $E(B\!-\!V)$ (equivalently $A_V$ and
$R_V=A_V/E(B\!-\!V)$, the parameterisation used in the text), the effective \vsini , and the distance
$D$. Blue (double-ringed) ellipses are quantities that follow deterministically via stellar physics:
the isochrones (dashed arrows) map mass and age to $(\teff,\logg,L)$; the stellar radius $R$ (needed for the 1D ATLAS and PHOENIX models) follows from
$(L,\teff)$ and is needed to set  the $10$\,pc flux normalisation of the
model spectrum $f_{\rm model}(\lambda)$ in the plane-parallel ATLAS and PHOENIX models
(Sect.~\ref{sec:models}). Model predictions and the observed data (grey) result in the three $\chi^2$
terms (red) of Eq.~\ref{eq:chi2}: extinction and distance enter only the XP term, the effective
\vsini\ only the BOSS term, and the parallax constrains $D$ directly. The constraint that the solutions need to be consistent with isochrones is
not enforced during the spectral fit but applied in post-processing (Sect.~\ref{sec:fitvsiso}):
the fit varies $(\teff,\logg)$ freely, and mass and age are constraint subsequently from
$(\teff,\logg,L_{\rm obs})$. The scheme repeats for each of the $N_\star$ stars; metallicity is held
at the solar value throughout.}
\end{figure*}

We set out to describe each star by a set of intrinsic stellar parameters (\Minit, \tauage , \vsini ), along with a set of extrinsic parameters: distance $d$, dust extinction (\AV \, and \RV). While acknowledging that the majority of stars are presumably in multiple systems, our analysis assumes that one component dominates the spectrum, as we only fit \emph{single-star models}\footnote{We are aware that for a significant fraction of massive stars this approximation does not hold: binary interactions, such as mass transfer, mergers, and stripping, reshape the evolution of a large fraction of the OB population \citep{Sana2012}. Our comparison against eclipsing-binary \emph{dynamical} masses (Sect.~\ref{sec:val_external}) nonetheless indicates that, for the large majority of systems, treating the spectra as single stars returns sensible parameters.}. Since the fundamental parameters \Minit\ and $\tauage$ are of course not directly observable in single stars, we translate them -- via isochrones -- to spectroscopic parameters \teff\ and \logg \, along with the luminosity \logLlam (\tauage , \Minit ).  Our goal is to devise a forward model that predicts the three rather different pieces of information we have for all stars: the continuum-normalized BOSS spectra in wavelength windows that encompass key diagnostic spectral lines; the precise spectrophotometry from the Gaia XP spectra; and the geometric distance estimate (parallax) that links the models' luminosities to the spectrophotometry once the extinction is accounted for. Forward-model purists might argue that we should find the best model parameters making \Minit\  \tauage\ fit parameters directly; here, we have opted to retain \teff\ and \logg \ as fitting parameters, as they predict the discriminating data features much more directly. This overall fitting approach is summarized and illustrated in Fig.~\ref{fig:schematic}.

\subsection{Spectral models}\label{sec:models}
Our forward model is based on a \emph{composite} grid of synthetic spectra that spans the full temperature
range of the initial sample, assembled from three families of models, each used where it is physically most
appropriate. Below $8\kK$ we adopt the PHOENIX-ACES LTE model atmospheres \citep{Husser2013}; from
$8$ to $15\kK$ the ATLAS9-based BOSZ LTE grid \citep{Bohlin2017}; and above $15\kK$ the PoWR expanding,
non-LTE, line-blanketed model atmospheres \citep{Hainich2019,Pauli2026}\footnote{Specifically the PoWR GAL-OB-Vd3 grid of Galactic OB-star models.}, which capture the winds and departures
from LTE that shape the spectra of the OB stars that are of primary interest. Each model is convolved
to the wavelength-dependent line-spread functions of \emph{Gaia} XP and of BOSS and resampled onto the
observed wavelength grids.

Placing the models on an \emph{absolute} flux scale lets the XP spectrophotometry and the
parallax constrain a luminosity. However, the three model families must be treated differently in this respect. The PoWR
models are computed for a specified stellar luminosity and radius and therefore already carry an
absolute flux. In contrast, the plane-parallel PHOENIX and ATLAS grids provide only the 
emergent flux \emph{per unit stellar surface area}, so they require a
radius to predict a total flux: we take $R(\teff,\logg)$ from the matched solar-metallicity MIST isochrone point (Sect.~\ref{sec:fitvsiso}) and scale
the surface flux according to the isochrone's radius.\footnote{PHOENIX and ATLAS tabulate the astrophysical surface flux
$F_\lambda$ (erg\,s$^{-1}$\,cm$^{-2}$\,\AA$^{-1}$ emergent per unit area of the stellar photosphere);
the flux received at distance $d$ is $F_\lambda\,(R/d)^2$. PoWR instead tabulates the flux already
referred to a distance for its native $(L,R)$.} All three families are in this way placed at a common
reference distance of $10$\,pc, and the single multiplicative flux factor carried in the fit then
rescales $10$\,pc to the stars' spectrophotometric distances.

The model building proceeds in two steps. First, assembling the set of spectral model grid points across the different model families; then devising an interpolator (to arbitrary \teff\, and \logg ), with an eye towards minimizing the impact of the initially rather coarse model grid. 

 The model families overlap in temperature, and we need stitch them into one contiguous grid using the union of their $(\teff,\logg)$ nodes. This leaves two ``seams'' in temperature:
PHOENIX-to-BOSZ near $8\kK$ and BOSZ-to-PoWR near $15\kK$. We chose to use the BOSZ grid down to $8\kK$, because this is well below the hot-star regime of interest here and so leaves the model seam in the regime that only affects the  cool contaminants. At the remaining seam the model families agree well: BOSZ and PoWR match to $\sim1.6\%$ in the continuum at $15\kK$. Where two model families provide a spectrum at the same (\teff,\logg)-node we average them.

Because their gravity sampling differs (PoWR is finer at $0.2$ dex, compared to $0.5$-dex for PHOENIX/ATLAS), we build the $\logg$ grid only from nodes that actually exist at each temperature, so that no empty $(\teff,\logg)$ cells are introduced. ATLAS and PoWR are further merged into a single
interpolation domain so that neither retains a hard grid wall at $15\kK$; this fills the $15$--$20\kK$,
high-$\logg$ region that neither family populates on its own and addresses a spurious pile-up of fits
at the seam.

As the second step, we then replace this discrete grid with a smooth interpolator, to obtain spectral model predictions at \emph{arbitrary} $(\teff,\logg)$. For each family we train a single thin-plate-spline emulator over the $(\log\teff,\logg)$
plane. In essence, a thin-plate spline is a two-dimensional generalization of a natural cubic spline \citep{Duchon1977} that results in a smooth, minimum-curvature interpolation that should be exact at the grid nodes. This provides a forward-model for all the observables (the XP spectrum and all BOSS windows) in one evaluation ($\sim\!0.4$\,ms). We found that this \emph{global} emulator across model families reduces the tendency that poorly constrained parameters (notably hot-star
$\logg$) remain ``stuck'' on their starting grid node, which had initially led to a ``banded structure'' in the \emph{Kiel-diagram} (Fig.~\ref{fig:kiel}). Since the grid is not rectangular (e.g., ATLAS has no low-gravity models at $15\kK$) an unconstrained interpolator can produce spurious extrapolations at the grids' edges. We therefore bound each emulator by the convex hull of its grid nodes, by adding a penalty term to $\chi^2$ that penalizes the fit if it ventures significantly beyond the (\teff , \logg)-hull.
The same continuous interpolation is applied to the grid-node \emph{luminosities} that set the models' absolute flux scale.

\subsection{Parameter Fitting}\label{sec:params}

To maximize the likelihood of the data for each star, i.e. the XP spectra, BOSS spectra and parallax, we have seven model parameters that can be varied:  effective temperature $\teff$ and surface gravity $\logg$; extinction $A_V$ and its shape $R_V$; the ``effective" projected rotation $\vsini$ (an \emph{effective} \vsini , reflecting a total line-broadening velocity that may arise from rotation, pulsations or macroturbulence; Sect.~\ref{sec:val_external}); the radial velocity\footnote{The co-added spectra are already in the rest frame; we merely allow a substantive $v_{\rm rad}$ adjustment in the rare cases where the data call for it.} $v_{\rm rad}$; and the distance to the star $d$ (see below).

Incorporating the stars' luminosity as an observational constraint in a ``spectral fit"  is crucial and subtle. We do it by solving for the distance $d$ that simultaneously satisfies the spectrophotometry and the parallax. To avoid double-counting the flux amplitude, we compare each model XP spectrum (defined at $10$~pc) to the data only after rescaling it to the observed integrated XP flux; that rescaling \emph{is} a spectrophotometric distance, which then enters the likelihood through the resulting $d$ and the $\varpi_{\rm pred}$ vs.\ $\varpi_{\rm obs}$ comparison. This $d$ is the sixth parameter. The atmospheric parameters $(\teff,\logg)$ enter through the emulator of Sect.~\ref{sec:models}.

Extinction is applied with the \citet{Gordon2023} (G23) law. Because G23 is linear in $1/R_V$ at every wavelength, we treat $R_V$ as continuous over its validity range $2.3\le R_V\le5.6$. Extinction enters only the XP term: the BOSS windows are continuum-normalised and hence insensitive to $A_V$ and $R_V$, so reddening is constrained entirely by the shape of the XP spectral energy distribution.
We fit for extinction in the range $0\le A_V\le12$~mag: a substantial number of hot stars in the inner disk genuinely exceed $A_V\sim 6$~mag, but still enter this optically selected sample. At low extinction ($A_V\lesssim1$~mag) the data carry essentially no information on the curve \emph{shape}, which may lead to pile-ups at the $R_V$ bounds; we therefore impose a weak Gaussian prior, $R_V=3.1\pm0.3$, which suppresses these pile-ups while leaving the well-constrained, strongly reddened stars essentially unaffected.

Each model is matched to the line-spread function (LSF) of both datasets. For XP the constraint comes from the \emph{broad} energy distribution, which is insensitive to LSF detail; for BOSS the Balmer and helium line widths constrain $\teff$, $\logg$ and especially $\vsini$, so the LSF matters. We therefore adopt the wavelength-dependent BOSS resolving power \citep{Smee2013}, which falls to $R\simeq1600$--$1900$ in the blue windows, and convolve each model on the coordinate $\varsigma=\int R(\lambda)\,{\rm d}\ln\lambda$, in which the broadening has fixed width; a constant-$R$ approximation would leave the lines too sharp and drive $\vsini$ spuriously high. Finally, $\vsini$ and $v_{\rm rad}$ are optimised jointly with the atmospheric parameters, the models being rotationally broadened with a pixel-integrated \citet{Gray2005} kernel and Doppler-shifted by $v_{\rm rad}$.

The best fit minimises the joint $\chi^2$ of the three data components,
\begin{equation}
\chi^2_{\rm tot} = \chi^2_{\rm XP} + \alpha\,\chi^2_{\rm BOSS} + \beta_\varpi\,\chi^2_\varpi,
\label{eq:chi2}
\end{equation}
where $\chi^2_{\rm XP}$ compares the XP flux to the reddened, distance-scaled model, $\chi^2_{\rm BOSS}$ the thirteen normalised BOSS windows, and $\chi^2_\varpi$ the inverse model distance to the observed parallax. Before entering $\chi^2_{\rm BOSS}$, individual pixels that are strongly deviant in the co-added BOSS spectra (cosmic-ray residuals and sky- or visit-combination artefacts) are sigma-clipped and masked, which significantly reduces the number of failed fits.

Because the three blocks contribute very different numbers of $\chi^2$-terms, we introduced $\alpha$ and $\beta_\varpi$ to set their relative weights. After some experimentation, we kept $\alpha=1$ , which gives equal weight to XP and BOSS, but \emph{up-weight} the single parallax term, $\beta_\varpi=10$. We did this because the parallax is a single \emph{direct} and in some sense exceptional datum on distance and luminosity of the star. Of course, the low-resolution spectroscopy also carries information about the luminosity, mostly via \logg . Yet, subtle systematics in the data (e.g. LSF) or models may lead to formally best fits at a spurious luminosity, and the $\gtrsim 100$ $\chi^2$ terms of the spectra may ``overwhelm" the parallax constraint in some cases. Re-weighting the blocks this way is the fractional (``power'') likelihood of ``safe-Bayes'' treatments of model misspecification \citep{Bissiri2016,Grunwald2017,Miller2019}. We chose not to downweigh the $\chi^2_{\rm XP}$ and $\alpha\,\chi^2_{\rm BOSS}$ terms to preserve their role as goodness of fit diagnostics, needed in the final sample cleaning. 

\begin{figure*}
\includegraphics[width=0.9\textwidth]{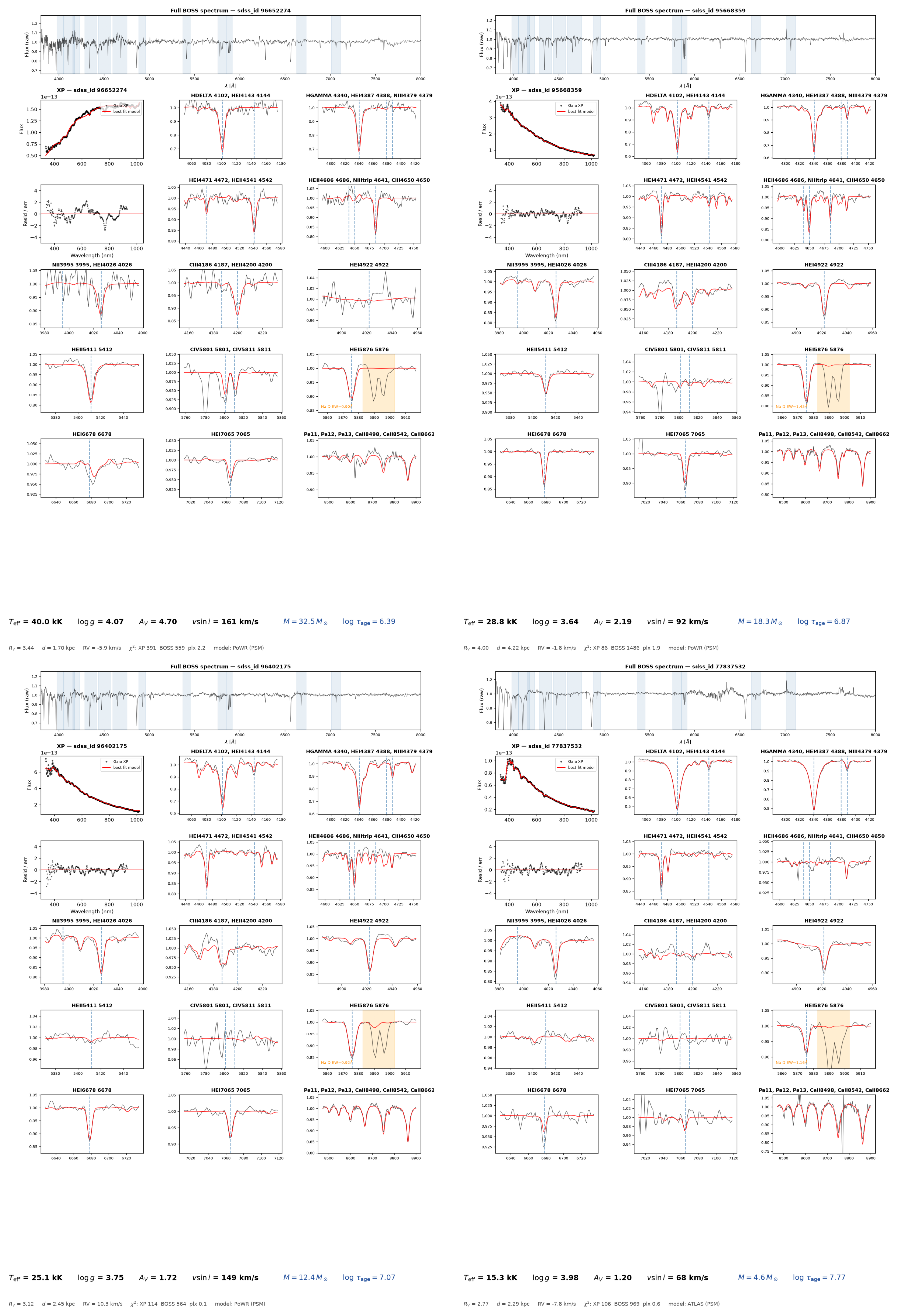}
\caption{\label{fig:examplefits}Example joint XP$+$BOSS fits for four stars spanning
$\teff\simeq15$--$40\kK$. Each panel shows the \emph{Gaia} XP spectral energy distribution and the
BOSS line windows (black) with the best-fit reddened model (red), and the fitted parameters.
\emph{Preliminary; final panels to be regenerated from the frozen catalog.}}
\end{figure*}

Three arguments imply that this re-weighting is robust and sensible, merely reducing the (already small) number of fits that have implausible best-fit luminosities and acknowledging the special role of the parallax measurement. \emph{Post facto}, the typical values of $\chi^2_\varpi\simeq1.4$ per d.o.f means that most fits are not forced onto the right parallax, while $\chi^2_{\rm BOSS}\simeq0.74$ shows the spectral fit quality has not been sacrificed. Re-fitting hot stars at $\beta_\varpi\in\{1,5,10,15\}$ leaves the median distances, masses and luminosities essentially unchanged for $\beta_\varpi\ge5$ (shifts $<\!1.5\%$, almost no star $>\!10\%$); only $\beta_\varpi=1$ differs, shortening median distances by $\sim\!6\%$ and moving over half the stars by $>\!10\%$. This shows that our chosen $\beta_\varpi=10$ sits on the ``flat part of the curve".
Fig.~\ref{fig:examplefits} shows representative joint fits across the temperature range.

\subsection{Fitted parameters vs. parameters inferred via isochrones}\label{sec:fitvsiso}
It is useful to distinguish here between parameters the data constrain directly and those that follow from our fundamental modelling assumptions (Fig.~\ref{fig:schematic}): we assume that all stars we model lie
\emph{near} the locus of solar-metallicity, stellar isochrones. This is equivalent to assuming that a single star dominates the spectrum which has evolved as a single star of a given initial mass and age. On an isochrone, (initial) mass and age are the \emph{underlying} parameters, in
that they determine $\teff$, $\logg$, the luminosity $L$, and the radius $R$; these quantities in turn
generate the predictions compared to the data: the radius and luminosity, together with the extinction
and an assumed distance, set the absolute XP spectrophotometry; $\teff$ and $\logg$, with $\vsini$ and
$v_{\rm rad}$, set the BOSS line windows; and the distance predicts the parallax.

In this formulation the quantity of central  interest, the stellar \emph{mass}, is the underlying
parameter, whereas the \emph{spectral luminosity} is what the data measure: the absolute XP flux fixes the
apparent bolometric flux, and the parallax converts it to $L$. The isochrone then maps that measured
luminosity, together with $\teff$, back to a mass. This is the reverse of the conventional procedure of
reading the mass directly off $(\teff,\logg)$. We have chosen this approach because the luminosity is measured
rather than inferred from the surface gravity, it provides the more informative constraint on the mass
wherever the parallax is well determined. Concretely, the single flux scale $s$ solved for at each model (Sect.~\ref{sec:params}; with the model fluxes defined at $10$\,pc) does double duty: it fixes the spectrophotometric distance, $d=10\,{\rm pc}/\sqrt{s}$, and, combined with the dereddened bolometric flux and the parallax, it fixes the measured luminosity $L$. This flux-derived luminosity, tabulated distance-independently as \texttt{logFbol} (Sect.~\ref{sec:products}), is the \emph{input} to the isochrone posterior, and is conceptually distinct from the isochrone-implied luminosity that accompanies the inferred mass and age.

One caveat bears mentioning:  we do not impose the
isochrone as a hard constraint: $\teff$ and $\logg$ are fitted freely (with a penalty being off isochrones), so a star that is \emph{not} a
single main-sequence object, e.g. an unrecognised binary, a stripped or subdwarf star, can be driven somewhat off the isochrone locus, which in turn may reveal its non-single-star nature. 

But more importantly, the mapping from
$(\teff,\logg)$ back to (mass, age) need not be single-valued; therefore, we carry a full posterior over
isochrone points rather than a point inversion. 
Given the fitted $(\teff,\logg)$ and the measured luminosity $L_{\rm obs}$, we compute a Bayesian
posterior over rotating MIST (v2.5, $v/v_{\rm crit}=0.4$) isochrone points \citep{Dotter2016,Choi2016}, weighted by a Kroupa initial mass function
\citep{Kroupa2001}, which is essentially identical to other common IMF forms above $\sim\!3\,\msun$, and by the evolutionary time spent at each point, from which we report the
posterior mass, age, luminosity and radius and their uncertainties. Where the parallax is informative,
the luminosity enters as data and dominates the mass estimate; where it is not, the estimate falls
back on the gravity. Adopting rotating isochrones turns out to matter little for the resulting masses but adjusts the overall age scale: relative to the solar-metallicity PARSEC grid \citep{Bressan2012}, considered for comparison, the rotating MIST isochrones leave the masses essentially unchanged ($+4\%$ in the median) while shifting all ages younger by a nearly uniform $0.08$\,dex ($\sim\!16\%$). Note that the uncertainties we report include systematic floors: $0.03$\,dex in $\log\teff$ ($\sim\!7\%$) and $0.10$\,dex in $\logg$, added in quadrature to the formal errors because the latter are photon-noise-only and too small to be used directly; the floors have been set so that the mass and age posteriors are not implausibly narrow.

To connect our model fits to more traditional approaches to estimating spectral types etc., we also measure equivalent widths of the
Balmer and helium lines.

\subsection{Interstellar absorption}\label{sec:byproducts}
 ($\lambda$8620)
Most hot stars lie at low Galactic latitudes, with much interstellar medium along the line of sight. This does not only cause reddening, but also spectral absorption features that are not intrinsic to the star, in particular the Na D doublet at 
$\lambda$5890 and a prominent DIB feature at $\lambda$8620. We exclude these regions from the fits, but then later on measure their equivalent widths and velocities.

\section{Results}\label{sec:results}
Our modelling returns formal parameter estimates in almost all cases, even for corrupt spectra, for very low
signal-to-noise ratio, and, quite frequently, for targeted stars that turned out to be not hot
($\teff<10\kK$). Before interpreting these parameters in detail we therefore ``clean'' the sample of the
sources whose formal labels cannot be usefully interpreted, or are not
hot stars. We do this in four steps: assessing the basic data quality, applying a qualitative spectroscopic hot-star criterion, restricting the sample to the Galactic disk, and finally restricting it to the stars whose spectra are part of SDSS DR20. We record the first three steps as \emph{flags} rather than hard cuts, so that the purity of
the sample can be tuned by the user; the last step defines the data-release scope of the published catalog. We then validate the resulting labels internally and externally
(Sect.~\ref{sec:validation}) and present their distributions (Sect.~\ref{sec:OBstar_population}).

\subsection{Defining the hot-star sample}\label{sec:sampledef}
Our fits formally converge for \Nfit\ of all SDSS-V-observed hot-star candidates. Five successive cuts, described in turn below and listed in Table~\ref{tab:gate}, take this set to the fiducial hot-star sample of \Nhot\ stars: (i) a \emph{data-quality} flag that removes corrupt or composite spectra; (ii) a \emph{signal-to-noise ratio} threshold (${\rm S/N}>20$); (iii) a \emph{model-independent spectroscopic} hot-star criterion based on the helium and metal lines; (iv) a geometric cut (distance, Magellanic-Cloud membership) that restricts the sample to the Galactic disk, leaving \NfidFull\ stars; and, last, (v) the data-release scope of this paper, which retains only stars whose co-added spectra are part of SDSS DR20.

\emph{Data quality.} Even after the deviant-pixel masking (Sect.~\ref{sec:params}), a
very small number of fits return an enormous BOSS $\chi^2$, $\chi^2_{\rm BOSS}>10^5$,
grossly separated from the $\lesssim10^3$ bulk; these flag irreparably corrupt or
mis-calibrated spectra (mostly in BOSS's red arm) and peculiar or composite objects that
the single-star model cannot represent. We set \texttt{boss\_bad} for these ($411$
stars, $0.1\%$). Next, we impose a signal-to-noise ratio
floor for our eventual fiducial sample of ${\rm S/N}>20$, which eliminates $\sim8\%$, leaving an analysis sample of $264{,}236$ stars. We found that below this S/N the individual absorption-line diagnostics that differentiate hot from cooler stars are no longer reliable.

\begin{figure}
\includegraphics[width=\columnwidth]{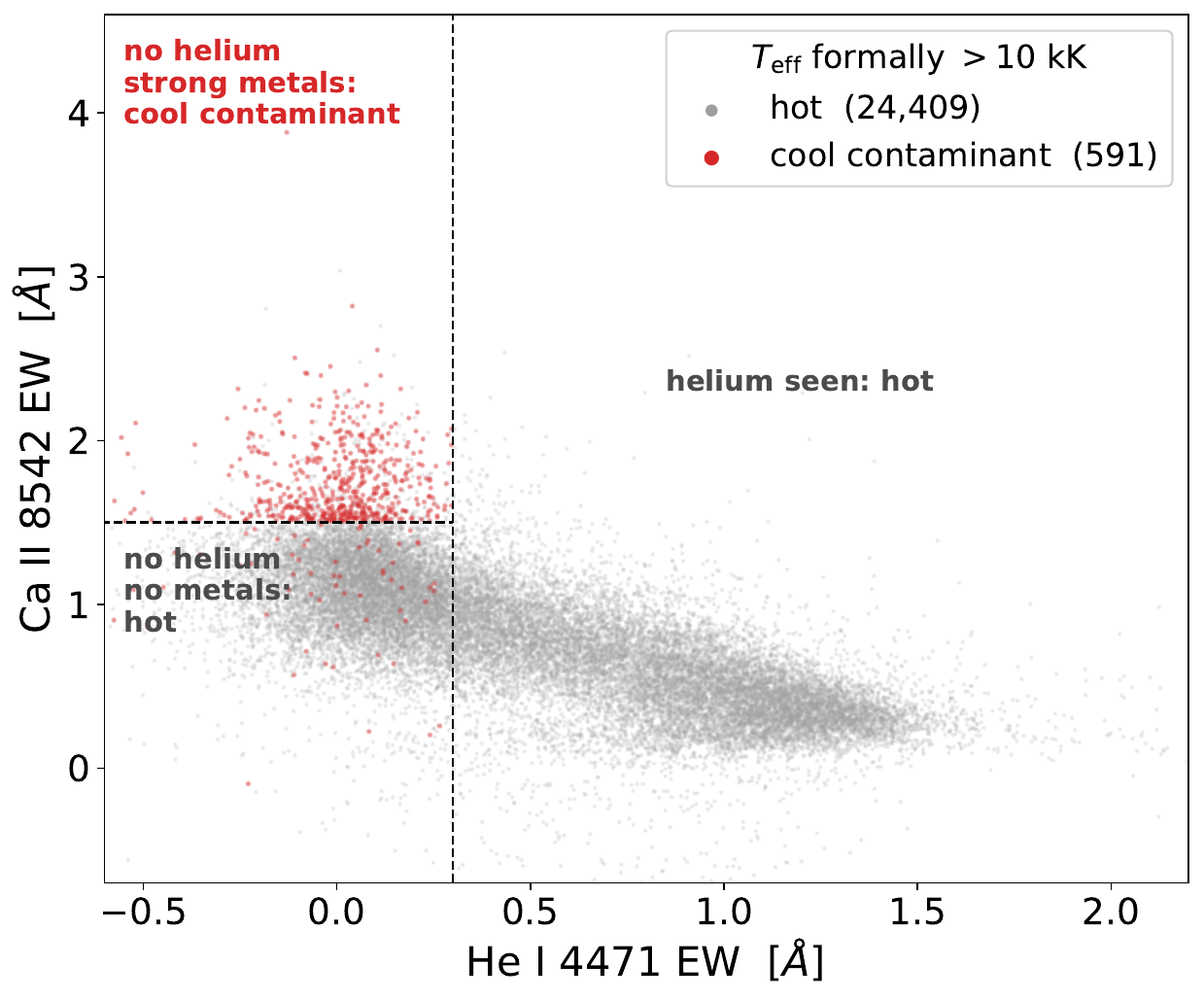}
\caption{\label{fig:gate}The hot-star criterion, for the $\teff>10\kK$ sample in the plane of two
decisive equivalent widths: He\,{\sc i}\,4471 (hot evidence) versus the extinction-surviving red
Ca\,{\sc ii}\,8542 line (cool evidence). Two classes are shown. Stars identified as hot are grey---those
showing helium, together with those showing neither helium nor metals (warm, fast-rotating stars whose
helium is rotationally smeared or intrinsically weak, kept as hot). Stars with no helium but distinct
metal lines are relabelled cool contaminants (red). For clarity, a random subset of $25{,}000$ stars with ${\rm S/N}>25$, drawn from the full (pre-DR20) sample, is shown.}
\end{figure}

\emph{A model-independent ``hot-star criterion".} Because temperature and extinction are strongly covariant in the SED, the Gaia XP
fit can misinterpret a reddened cool star as a spurious hot, highly reddened solution: the
spectrophotometry can ``outvote'' the spectral line windows in the BOSS spectra, contaminating any hot-star selection. We
address this not by a cut on the fits' $\teff$ values but by \emph{positive} identification of diagnostic lines in the BOSS
windows. A star is labelled hot if it shows genuine helium (He\,{\sc i}\,4471/4026, or He\,{\sc ii}\,4542
above $30\kK$, where He\,{\sc i} has ionised away). (Only) if the spectra do not show clear helium lines, we examine possible metal-lines: a spectrum without clear Helium lines is labelled a cool contaminant if it
shows the metal lines that hot stars lack, most often Ca\,{\sc ii}\,K, with the red Ca\,{\sc ii} triplet
as an extinction-insensitive fallback for the most reddened stars (see Fig.~\ref{fig:paschen}). Crucially,
we do \emph{not} treat the absence of helium alone as ``cool'': e.g. a very hot fast rotator might have its helium lines
rotationally smeared below detection; so a spectrum with neither helium nor metal lines discernable is kept as hot. 
We quantify ``discernable'' by equivalent width (EW) criteria:  we treat a line as present when its EW exceeds a line-specific threshold, namely $0.3$\,\AA\ for He\,{\sc i}\,4471, $0.4$\,\AA\ for He\,{\sc i}\,4026 and $0.3$\,\AA\ for He\,{\sc ii}\,4542
on the hot side, and $2.0$\,\AA\ for Ca\,{\sc ii}\,K (or $1.5$\,\AA\ for the red Ca\,{\sc ii}\,8542 line)
on the cool side. Fig.~\ref{fig:gate} shows the resulting decision in the plane of two decisive
widths, He\,{\sc i}\,4471 (hot) versus the extinction-surviving red Ca\,{\sc ii}\,8542 (cool). For the $126{,}264$ fits with $\teff>10\kK$ that pass the quality and S/N cuts (i.e.\ before the disk and DR20 restrictions), this logic yields four categories:
\texttt{confirmed\_hot} (helium seen; $95{,}443$), \texttt{hot\_no\_metals} (helium smeared or
intrinsically weak; $27{,}261$), \texttt{cool\_contaminant} (metals seen, implying spuriously-hot fits,
are relabelled; $3{,}366$), and a small residual (\texttt{no\_boss}; $194$) that could not be classified

\emph{The fiducial SDSS-V hot-star sample.} To create a fiducial sample of hot stars, we require all criteria of Table~\ref{tab:gate} to be satisfied: acceptable $\chi^2$, and S/N$>20$. As this is intended to be a Galactic disk catalog, we also exclude the $\sim3{,}500$ Magellanic
Cloud members (identified by position, proper motion and parallax; fitted with a literature distance prior rather than the parallax) and the small tail of stars with spectrophotometric distances  $>20$\,kpc, in most cases associated with an ill-constrained parallax.  Within this sample, of course the helium-lines-seen tier is the
purest. For maximum purity one can restrict a sample to the helium-seen tier and to $\teff>13\kK$, where the criterion is essentially clean.
Warmer A-type
($7$--$10\kK$) and cooler stars are tabulated separately (Appendices~\ref{sec:appA},~\ref{sec:appB}).
The fiducial disk sample numbers \Nfiducial\
stars, with Table~\ref{tab:gate} listing the surviving counts at each successive step. Explicitly, it is the set of stars satisfying \texttt{hot\_confirmed} (which enforces $\teff>10\kK$) $\wedge$ $\lnot$\texttt{boss\_bad} $\wedge$ ${\rm S/N}>20$ $\wedge$ $d<20$\,kpc $\wedge$ $\lnot$\texttt{in\_MCs} $\wedge$ \texttt{majority\_before}. The last flag implements the data-release scope of this paper, limited to stars with the majority of epochs observed within SDSS DR20 (i.e. before MJD 60708; 2025 February 2). In this work we use the larger set only for \emph{validating} the labels (Sect.~\ref{sec:validation}), but all population results and the published catalog refer to the \Nfiducial\ DR20 stars. We restrict the subsequent discussion and analysis to this fiducial sample.

\begin{figure*}
\includegraphics[width=\textwidth]{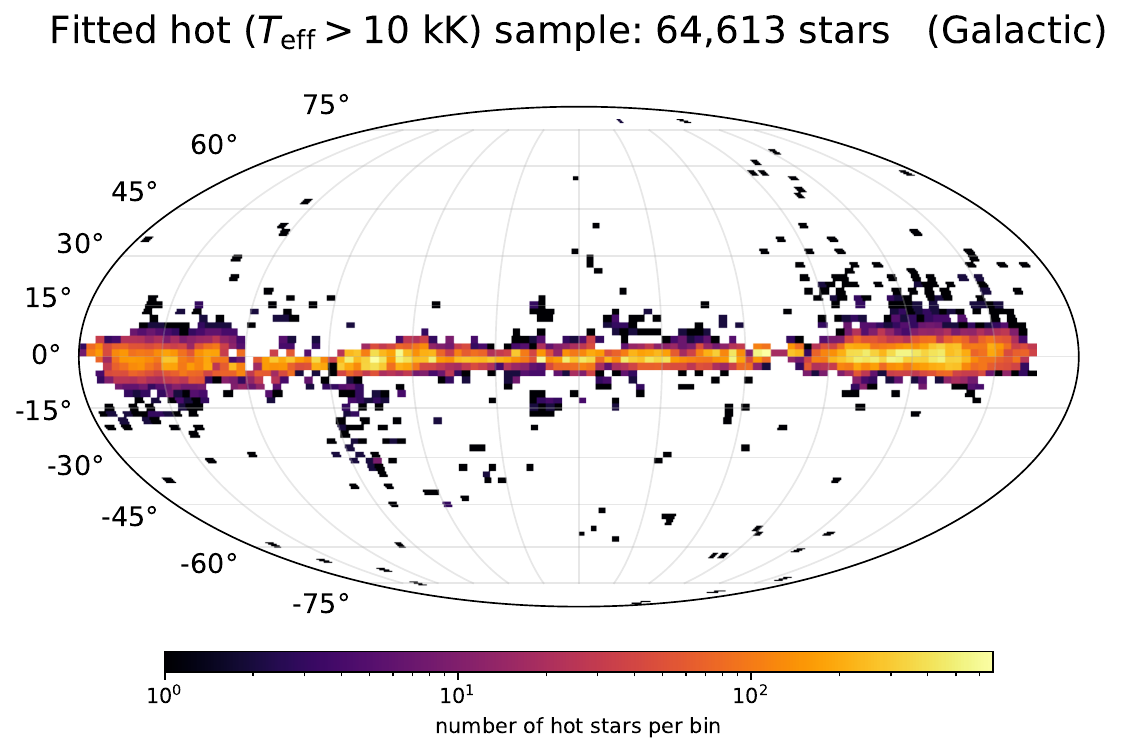}
\caption{\label{fig:footprint}Galactic distribution of the fiducial hot ($\teff>10\kK$) sample
($\Nhot$ stars), shown as a number-density map in Galactic coordinates. The sample traces the young, low-latitude disk; the longitude patchiness
reflects the SDSS-V field coverage.}
\end{figure*}

Fig.~\ref{fig:footprint} shows the sky distribution of this sample: all-sky, but strongly concentrated toward the Galactic plane and patchy in longitude, reflecting the operational fiber assignment
(Sect.~\ref{sec:targets}).

{
\setlength{\tabcolsep}{2pt}
\begin{deluxetable}{lrr}
\tablecaption{From the fitted sample to the fiducial hot-star sample. Each row applies one flag,
cumulatively (i.e.\ in addition to all rows above it), and lists the number of stars surviving; the
final row is the fiducial sample used throughout.\label{tab:gate}}
\tablehead{\colhead{Selection of Fiducial Sample} & \colhead{$N$} & \colhead{removed}}
\startdata
Fitted (parameters returned) & $287{,}435$ & \nodata \\
\quad good BOSS $\chi^2$ (\texttt{!boss\_bad}) & $287{,}024$ & $411$ \\
\quad ${\rm S/N}>20$ & $264{,}236$ & $22{,}788$ \\
\quad hot\_confirmed (\texttt{helium/metal lines}) & $122{,}704$ & $141{,}532$ \\
\quad Galactic disk ($d<20$\,kpc, non-MC) & $119{,}121$ & $3{,}583$ \\
\quad DR20 coadds (\texttt{majority\_before}) & $64{,}613$ & $54{,}508$ \\
\enddata
\tablecomments{The hot-star step removes all non-hot categories together: cool ($\teff<10\kK$;
$137{,}972$), \texttt{cool\_contaminant} ($3{,}366$), and \texttt{no\_boss} ($194$). The final row
restricts to stars whose co-added spectra are published in DR20, i.e.\ at least half of their epochs
precede the DR20 cutoff (MJD 60708, 2025 February 2); the penultimate row ($\NfidFull$ stars) is the
same selection without the DR20 restriction and is used for the label-validation tests. All counts in this table refer to the revised fits (Sect.~\ref{sec:models}--\ref{sec:params}) and the DR20 scope.}
\end{deluxetable}
}

\subsection{Validation of the stellar labels}\label{sec:validation}
We validate the fitted stellar labels in the fiducial catalog along two lines. \emph{Externally}, we compare against independent measurements
of the same stars from surveys with different instruments and analysis codes: dynamical masses from
eclipsing binaries, and effective temperatures from the IACOB and LAMOST surveys; these can test our
absolute scale. \emph{Internally}, we compare against the SDSS-V BOSSNet pipeline, which analyses the
\emph{identical} BOSS spectra with an independent (neural-network) method; this removes any
data-difference and isolates methodological differences in the stellar label estimates.

\subsubsection{External validation}\label{sec:val_external}
(All validation statistics in this and the following subsection have been recomputed on the revised catalog; the changes are small. The temperature and $\vsini$ comparisons use the full-sample fiducial set of \NfidFull\ stars, i.e.\ without the DR20 restriction.)

\begin{figure*}
\gridline{\fig{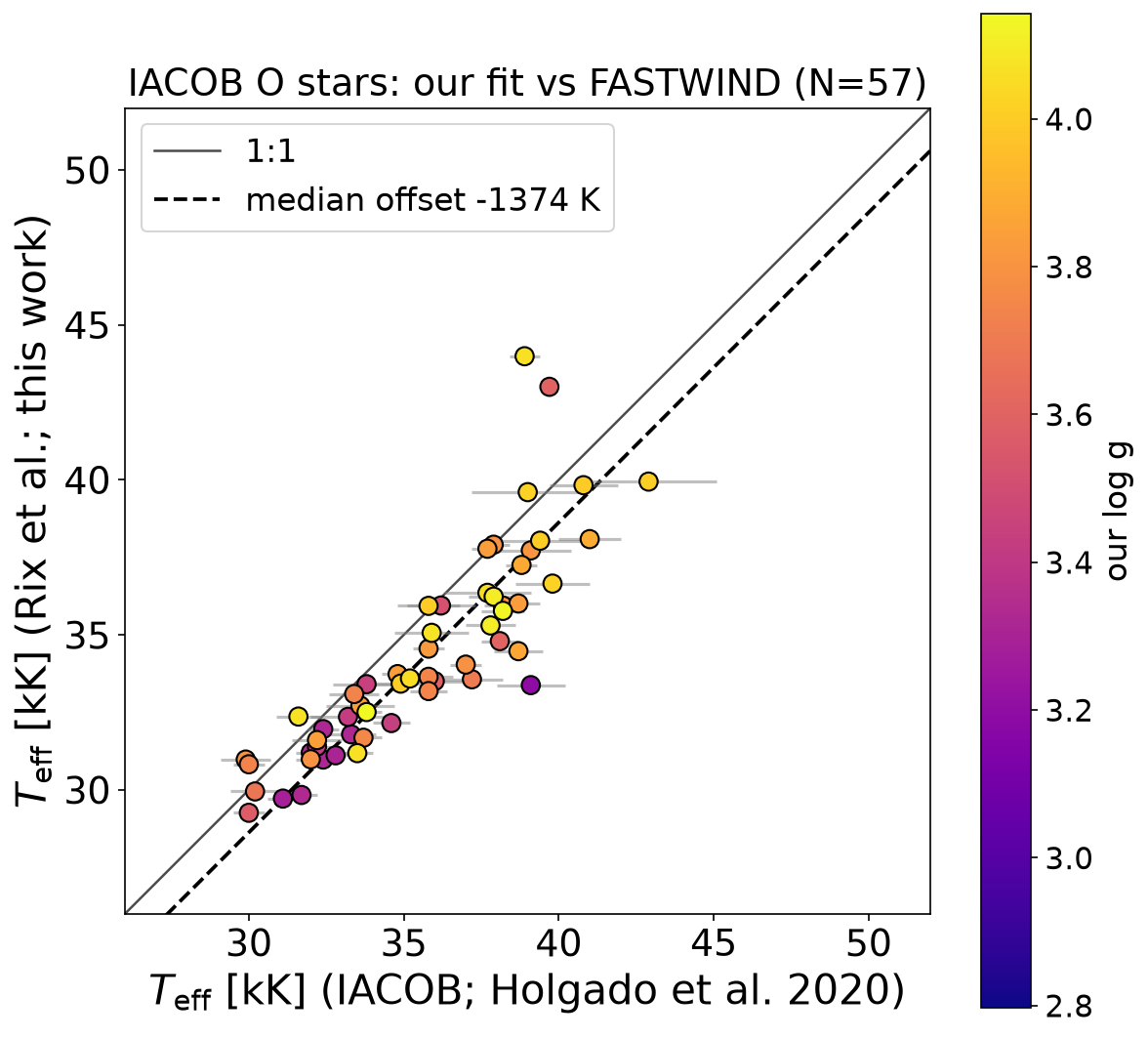}{0.5\textwidth}{(a) IACOB (FASTWIND O stars)}
          \fig{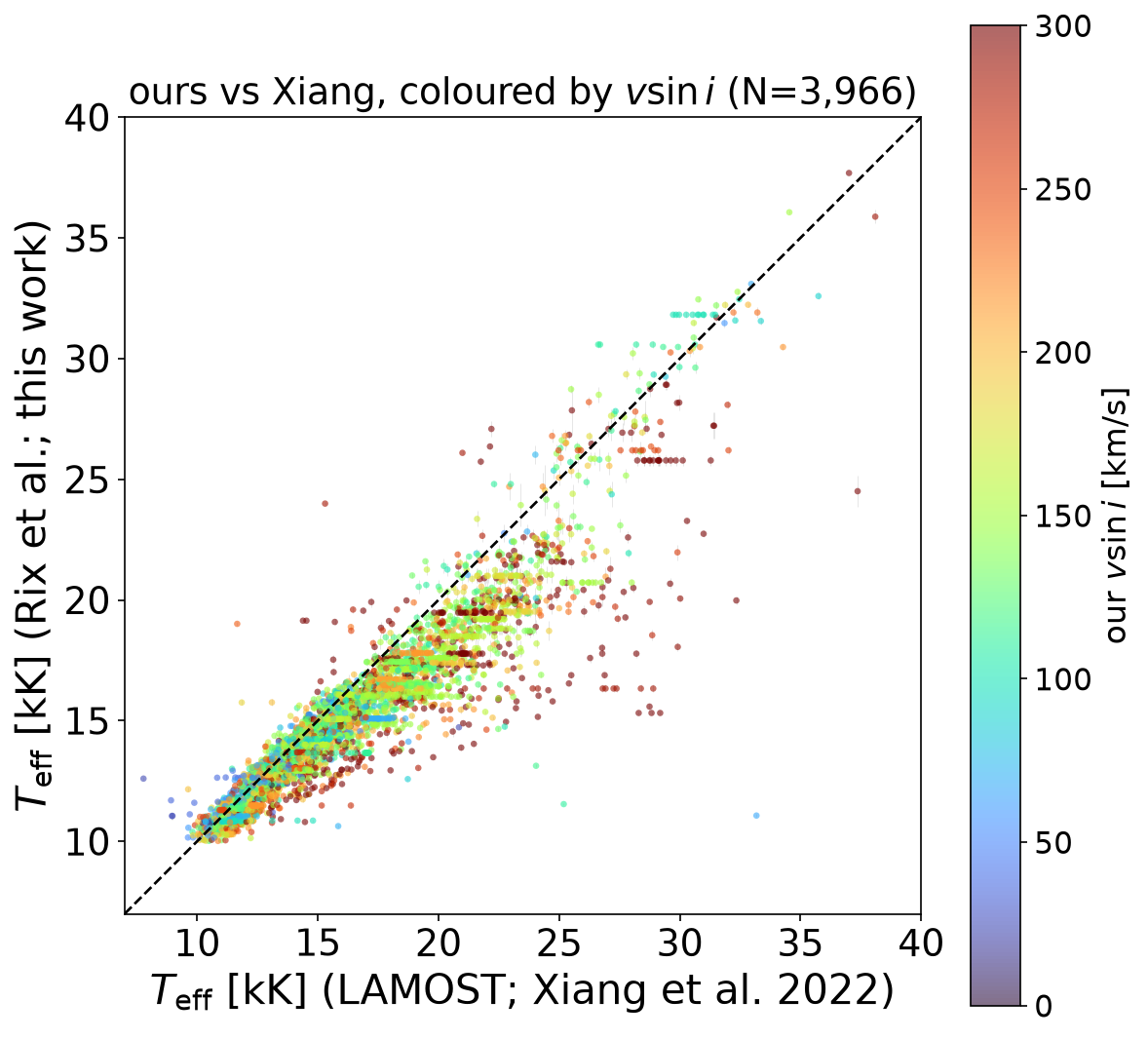}{0.5\textwidth}{(b) LAMOST (Xiang et al.)}}
\caption{\label{fig:valteff}Effective-temperature validation. (a) Our $\teff$ versus the IACOB
FASTWIND scale for 57 O stars: a $-3.8\%$ ($-1.4\kK$) offset (dashed), the known code-dependent
O-star scale difference. (b) Our $\teff$ versus LAMOST at scale ($N=3{,}966$), coloured by $\vsini$;
the two agree at the cool and hot ends, while our values are lower by $\sim\!7\%$ at 15--20\,kK and $\sim\!10$--$13\%$ at 20--30\,kK, presumably reflecting non-LTE effects absent from the LTE LAMOST analysis.}
\end{figure*}

\textbf{Effective temperatures.} The arguably most obvious stellar label to validate is  $\teff$, which we do by comparing against against two independent scales
(Fig.~\ref{fig:valteff}). First, against the IACOB survey \citep{Holgado2018,Holgado2020}, whose O-star temperatures come
from \textsc{fastwind} model-atmosphere fits: we find a small offset of $-3.8\%$ ($-1.4\kK$) over 57
stars (star-to-star scatter, as median absolute deviation, $2.5\%$, i.e.\ $\sim\!0.9\kK$; standard error on the median $0.7\%$), which seems consistent with expected code-dependent differences among O-star $\teff$ scales \citep[e.g.,][]{Sander2024_XShootU_IV}, and comparable to the systematic spread \emph{between} such analyses. 
The second comparison is against LAMOST hot-star parameters
\citep{Xiang2022}, where there are $N=3{,}966$ stars in common: the two scales agree to within $\sim\!3\%$ below $15\kK$ and above $30\kK$, while our temperatures are lower by $\sim\!7\%$ at $15$--$20\kK$ and by $\sim\!10$--$13\%$ at $20$--$30\kK$. This offset, largest at $20$--$30\kK$, is plausibly attributed to non-LTE effects, which are included in the PoWR models at these temperatures, but not in the LTE grids of \citet{Xiang2022}, rather than by rotation. Together, these place our temperatures on the established scale to within modest inter-survey systematics. It should be noted that our temperature scale may be the most robust, as our fits incorporate both individual diagnostic absoprtion lines and spectrophotometry.

\textbf{Effective line broadening ($\vsini$)}
Several effects broaden the absorption lines of these hot stars. The instrumental
line-spread function is modelled explicitly (Sect.~\ref{sec:params}). Beyond it there are two effects
in hot stars the can greatly broaden the lines. The first is stellar rotation that may reach $\ge 300$~km/s. The second is related to stellar pulsations. These may manifest themselves as macroturbulent 
fields of high velocity amplitude across much of
the OB domain, generated by sub-surface convection and stochastically excited gravity 
waves \citep[pulsations that are damped below the photosphere, yet collectively perturb it;][]{Aerts2009,Cantiello2009,AertsRogers2015,Grassitelli2015,SimonDiaz2017}.
Space photometry has shown the corresponding stochastic variability to be ubiquitous
among OB stars \citep{Bowman2019}. Our model fits a single rotational kernel, so the
quantity we measure is an \emph{effective} $\vsini$: the total line broadening, expressed
as a rotation velocity. Rotation is the dominant broadener certainly above
$\sim\!100\kms$; below that, macroturbulence contributes tens of $\kms$, and the fact
that $\vsini\ge0$ adds a one-sided noise bias of order the per-star uncertainty
($\sim\!40$--$50\kms$). For slow rotators, the effective $\vsini$ should therefore be
read as an upper envelope on the true rotation rate.

\begin{figure}
\includegraphics[width=\columnwidth]{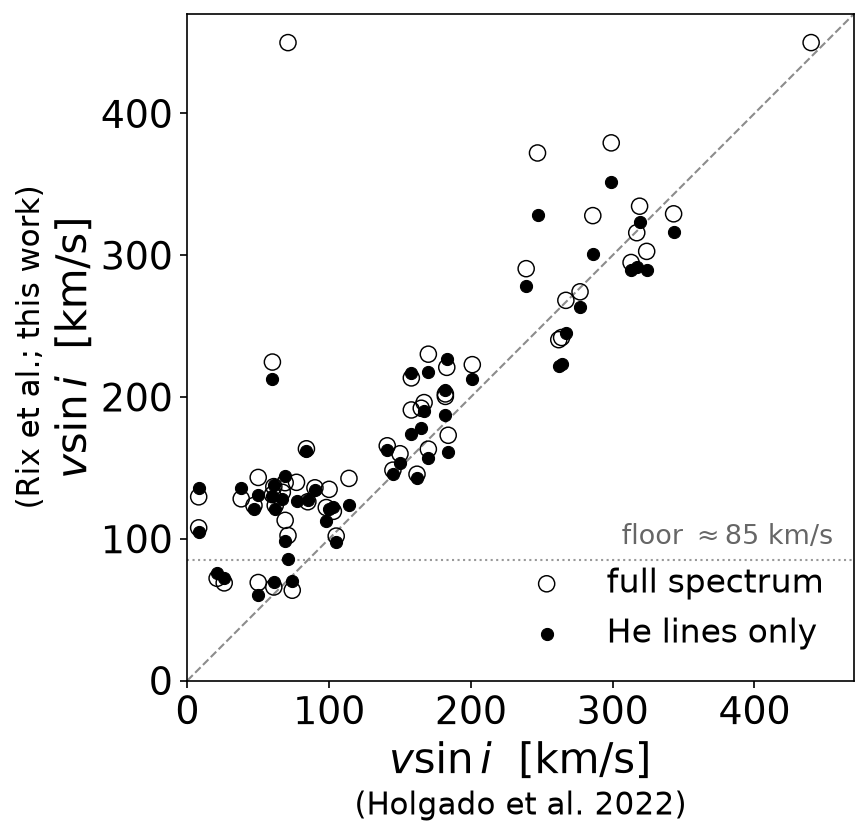}
\caption{\label{fig:valvsini}Projected rotational velocity, $\vsini$, against the IACOB Galactic O-star sample
\citep{Holgado2022}, for $57$ stars in common. Filled circles use our Helium-lines-only
$\vsini$, open circles the full-spectrum value; the dashed line is $1{:}1$. Above $\sim\!100\kms$ the
labels track the literature ($r=0.92$); below, they saturate at a floor of $\approx\!85\kms$ (dotted).
The floor is in \emph{our} values (the ordinate), set by a one-sided noise bias on slow rotators rather
than by the instrumental resolution; it is a statistical floor, not a hard cut, so the few stars that
fall below it while matching the literature $\vsini$ are expected.}
\end{figure}

Given BOSS's resolution of $R\!\approx\!1800$ in the blue diagnostic windows (an
instrumental FWHM of $\sim\!170\kms$), the line broadening needs particular empirical validation. We
compare against the IACOB sample of Galactic O~stars \citep{Holgado2022}, matching $57$
stars to our fiducial sample (Fig.~\ref{fig:valvsini}). Above $\sim\!100\kms$ the
agreement is excellent: a median offset of $+16\kms$ for the full-spectrum value and
$+5\kms$ for the He-lines-only estimate. Below $\sim\!100\kms$ our values saturate at a
floor of $\approx\!85\kms$, as the combination of macroturbulent broadening and the
one-sided noise bias predicts. That this floor is not an artefact of a mis-matched
instrumental profile is shown by the cool giant interlopers in the initial sample
($4.5<\teff<5.5\kK$): individually they are recovered down to $\vsini\approx14\kms$ (10th percentile), while
their population median of $\approx\!55\kms$, for stars that must be intrinsically slow
rotators, calibrates the noise contribution to the floor.
For stars well above 10 kK, the $\vsini$ from only He lines tracks the literature more
closely, and is preferred where available.

\begin{figure}
\includegraphics[width=\columnwidth]{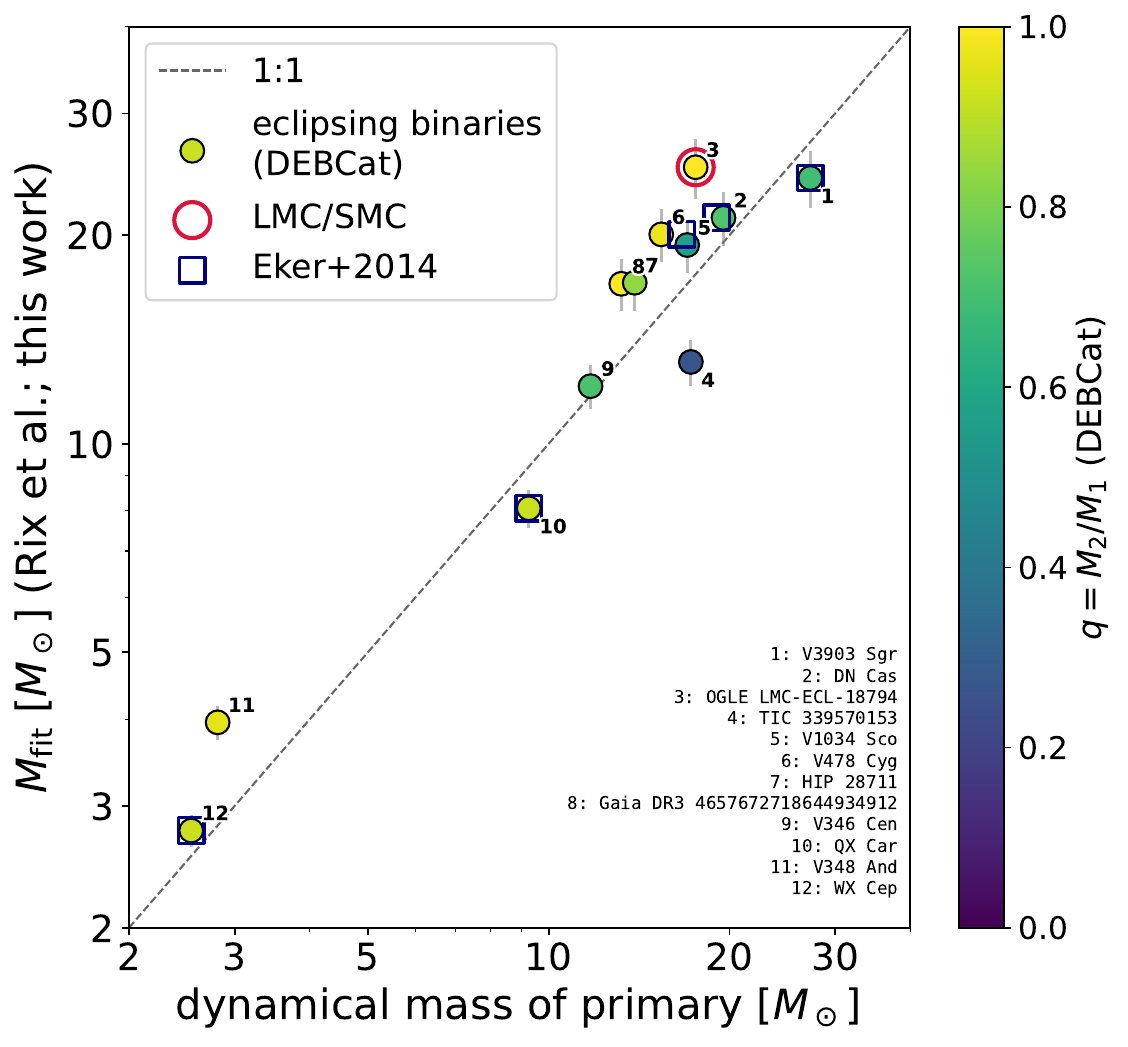}
\caption{\label{fig:valmass}Our single-star masses versus eclipsing-binary primary \emph{dynamical}
masses (DEBCat + Eker), over $\sim2$--$27\,\msun$; the median ratio is $\approx1.09$ for the Galactic
systems. Points are numbered by descending mass (key, lower right).}
\end{figure}

\textbf{Dynamical masses.} Part of our fits are isochrone-implied stellar masses. Here we can do a stringent comparison with an albeit-small sample of 
eclipsing-binary primaries, whose masses are \emph{dynamical} and model-independent. Cross-matching to
the DEBCat compilation \citep{Eker2014} and to \citet{Eker2018} yields a sample spanning $\sim2$--$27\,\msun$
(Fig.~\ref{fig:valmass}); the median ratio of our single-star mass to the dynamical mass is
$\approx1.09$ for the Galactic systems, i.e.\ a median offset of $\sim\!9\%$ with a star-to-star scatter of $\sim\!0.1$\,dex and no significant trend with
mass. We deem this a non-trivial success: our masses are inferred through the isochrones from the measured
luminosity and $\teff$ (Sect.~\ref{sec:fitvsiso}), yet they agree well with the dynamical measurements. This comparison also illustrated that for the vast majority of binary systems treating our spectra as single stars leads to sensible parameters. We additionally checked our (evolutionary) masses against \emph{spectroscopic} masses computed from the fitted $\logg$ and the radius implied by $L$ and $\teff$: these run $\sim\!10\%$ below the evolutionary values and the offset grows toward lower gravities, consistent with the long-known spectroscopic-versus-evolutionary mass discrepancy \citep{Herrero1992}.

\subsubsection{Cross-check with the SDSS-V BOSSNet pipeline}\label{sec:val_bossnet}
\begin{figure}
\includegraphics[width=\columnwidth]{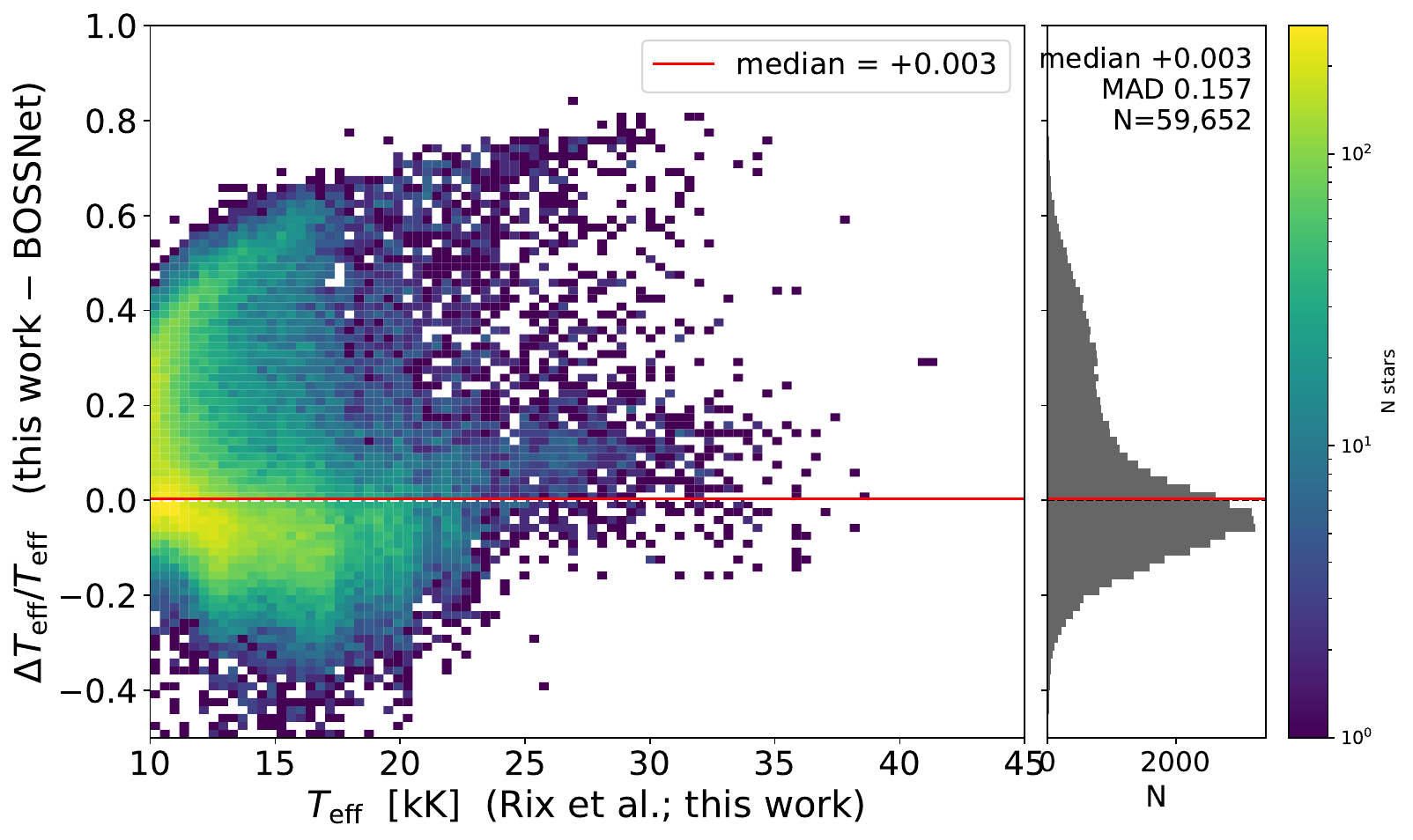}
\caption{\label{fig:bossnet} Cross-check against the SDSS-V BOSSNet pipeline (DR22,
astra~0.8.1) on the \emph{identical} spectra, for our He-confirmed hot sample
(\texttt{hot\_confirmed}) passing the quality cut: the fractional residual
$\Delta\teff/\teff=($ours$-$BOSSNet$)/$ours versus our $\teff$, with its marginal distribution
(right). The median fractional offset is $+0.003$ (MAD $0.16$). The positive tail at
$\teff\lesssim15\kK$ is BOSSNet underestimating warm stars, as it apparently shifts late-B/A0 objects toward
$\sim7\kK$ regardless of their Balmer-line strength (quantified in the text above).}
\end{figure}

BOSSNet \citep{Sizemore2024} is the data-driven neural-network stellar-parameter pipeline run within
the SDSS-V/Astra framework; it ingests the same BOSS spectra we do (we use DR22, astra~0.8.1). Matching to our He-confirmed hot
sample that passes the quality cut ($N=59{,}652$; Sect.~\ref{sec:sampledef}), the \emph{average}
agreement at first seems excellent: a median fractional temperature residual of $+0.3\%$ and a median $\logg$ offset
of $-0.01$\,dex, with a $\logg$ scatter of $0.26$\,dex typical of spectroscopic gravities
(Fig.~\ref{fig:bossnet}). 

This comparison of averages, however, hides  several \emph{regime-dependent serious systematics}, as Fig.~\ref{fig:bossnet} shows. First,
BOSSNet systematically shifts a subset of hot stars to cool temperatures: of our confirmed-hot stars at $10$--$13\kK$, $36\%$ are assigned $\teff<9\kK$ by BOSSNet, an incidence that falls monotonically with
temperature to $\sim2\%$ above $30\kK$. As we detect significant He\,{\sc i} absorption, we
presume that this is mostly a shortcoming of the neural net, not of our labels. Second,
the median offset changes sign with temperature: BOSSNet reads $\sim7\%$ low at $10$--$13\kK$, a few
percent \emph{high} at $15$--$20\kK$, and $\sim12\%$ low again above $25\kK$. Overlaid on these trends
is a persistent $\sim17\%$ fraction of outliers ($|\Delta\teff/\teff|>0.3$) at
\emph{all} temperatures. We conclude from this that BOSSNet is doing well in determining \teff\, for many hot stars, but has important failure modes that our modeling should remedy.

\begin{figure}
\includegraphics[width=\columnwidth]{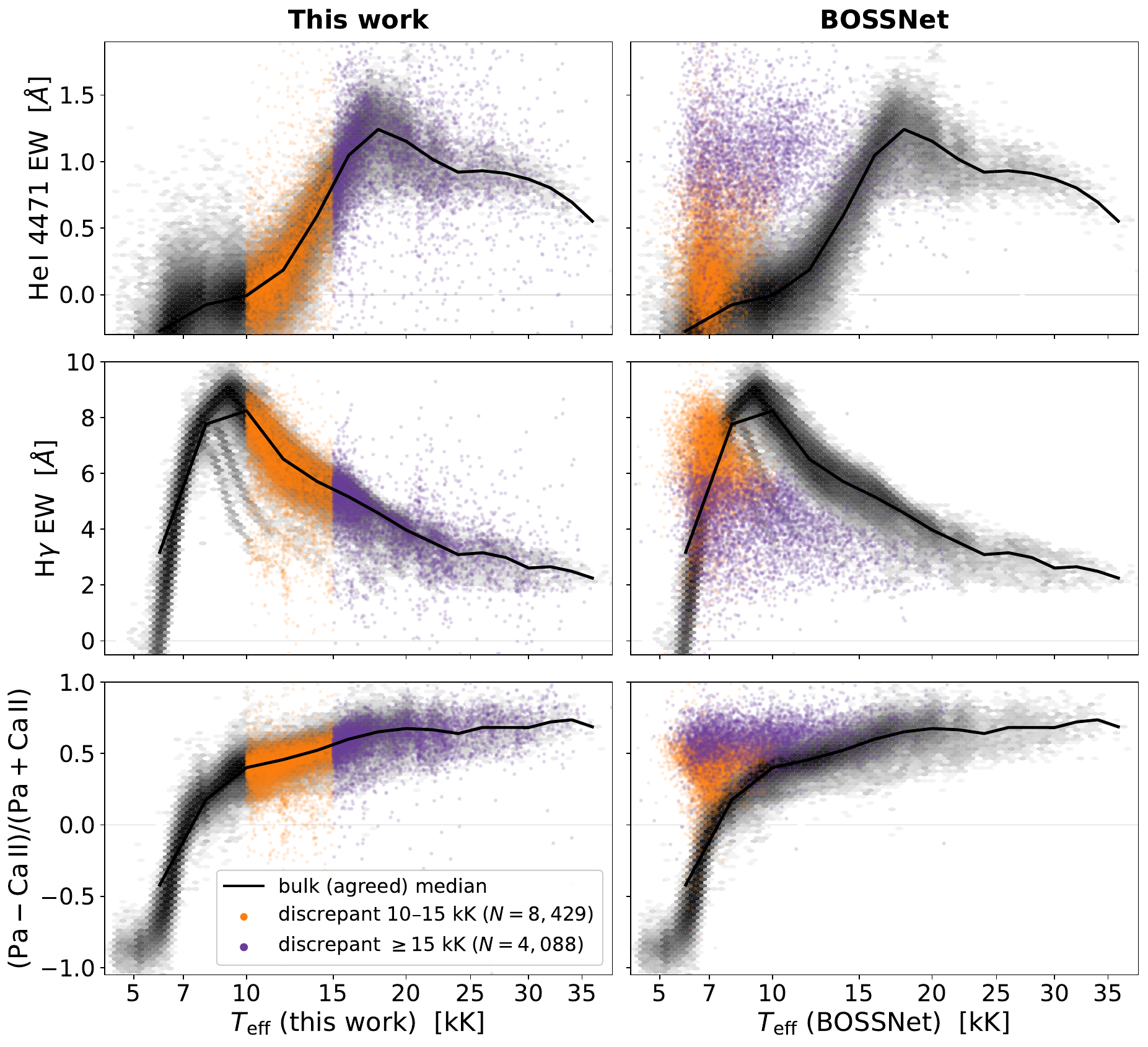}
\caption{\label{fig:bossnetew}Adjudicating the outliers in the BOSSNet comparison with equivalent widths
measured in this work. Rows: He\,{\sc i}\,4471,
H$\gamma$, and a far-red index
$(\mathrm{Pa}-\mathrm{Ca\,II})/(\mathrm{Pa}+\mathrm{Ca\,II})$ sensitive in the $6$--$12\kK$ regime; each is plotted against our \teff\
(left) and BOSSNet's (right); grey is the log-density of the \emph{agreed} sample
($|\Delta\teff/\teff|\le0.3$) with its running median (black). The outliers
are overplotted, in orange for our $\teff=10$--$15\kK$, in purple
for $\ge15\kK$. A star joins the grey locus at its \emph{correct} temperature, showing that EW diagnostic mostly support our $\teff$ from global fits as correct. The $x$-axis
is scaled as $\sqrt{\teff}$, highlight the $7$--$20\kK$ range.}
\end{figure}

To adjudicate these discrepancies we turn to the equivalent widths (EWs) of individual diagnostic lines,
measured directly in our BOSS spectra and hence \emph{independent} of either parameter pipeline. The
logic is simple: the strength of a temperature-sensitive line is a near-monotonic function of \teff\
over a wide range, so a star's measured EW should fall on the locus traced by the stars on which the two
pipelines agree; and it should do so at its \emph{correct} temperature. Fig.~\ref{fig:bossnetew}
plots three such diagnostics, namely He\,{\sc i}\,4471, H$\gamma$, and the reddening-immune far-red index
$(\mathrm{Pa}-\mathrm{Ca\,II})/(\mathrm{Pa}+\mathrm{Ca\,II})$ (Sect.~\ref{sec:products}), against our
\teff\ (left) and against BOSSNet's (right), with the agreed-sample locus in grey. The $\sim$$14{,}000$ outliers ($|\Delta\teff/\teff|>0.3$, where BOSSNet reads cooler) are overplotted, split by
our \teff.

The figure provides a consistent attribution of these discrepancies. Discrepancies among genuinely hot  (our $\teff\ge15\kK$; $N\approx4{,}100$) exhibit strong He,{\sc i} absorption that would be physically impossible at the $7$--$9\kK$ BOSSNet assigns them: $93\%$ show He,{\sc i},4471$,>0.3$,\AA. Discrepancies for cooler systems (our $10$--$15\kK$; $N\approx8{,}400$) behave the same way, though less extreme. This is best illustrated by the far-red Paschen/Ca,{\sc ii} index (monotonic in \teff) which places the vast majority of these systems near our $\sim$$12\kK$ rather than BOSSNet's $\sim$$7\kK$, as they exhibit weak Ca,{\sc ii},K ($\sim$$0.6$\,\AA, versus $>2$,\AA\ for true A/F stars). So, in both regimes the discrepancies should be mostly attributed to the imperfect temperature estimates of BOSSNet. At BOSS resolution there are at least two plausible channels that can lead to such mis-attribution: with rotationally broadened lines, H$\epsilon$ blends into Ca\,{\sc ii}\,H (3968\,\AA), so hydrogen absorption can be read as Ca\,{\sc ii} strength and hence as a cooler type; and Fe\,{\sc ii}\,5169 (which can be present in hot stars, including as a wind line) falls within the Mg\,{\sc i}\,b triplet (5167/5172/5183\,\AA), so iron absorption from a hot star can masquerade as the Mg\,{\sc i}\,b of a cool one. This exercise also shows the value of individual EWs of particular diagnostic absorption lines (Sect.~\ref{sec:products}): they are a compact, model-independent handle on the stellar parameters, diagnostic in the regimes where the global fits can be fragile.

\subsection{Properties of the fiducial SDSS-V OB sample}\label{sec:OBstar_population}
We now turn to the distribution of stellar labels across the fiducial sample and across the Galaxy. We defer any discussion of the sample selection function and completeness to Sect.~\ref{sec:selfunc}.

\begin{figure*}
\includegraphics[width=\textwidth]{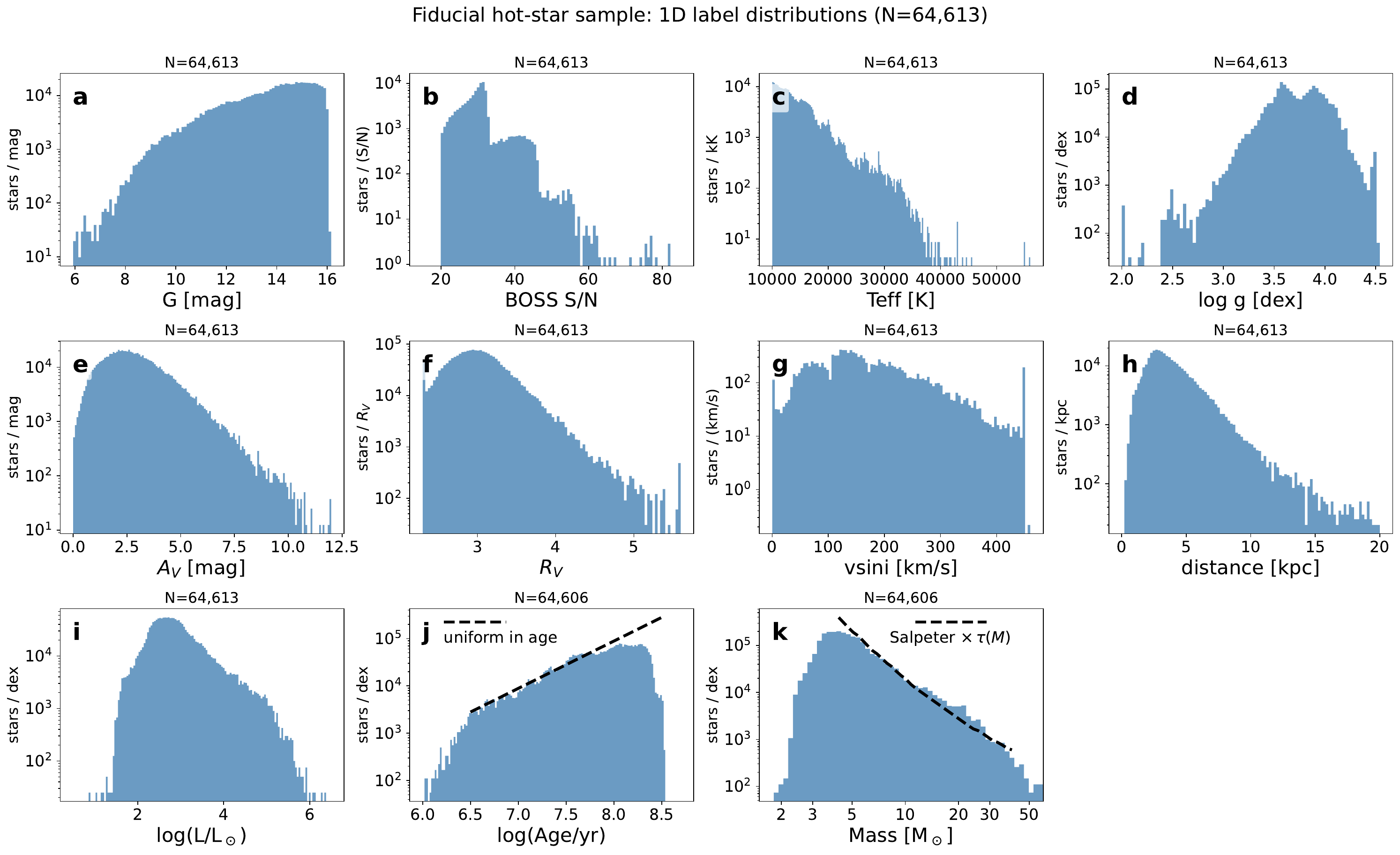}
\caption{\label{fig:1Dhistograms}Distributions of the catalog quantities for the fiducial hot-star sample: sub-panel (a) shows the magnitude distribution, with targets $G<13$ enabled by offsetting the fibers from the targets; the spectra's mean S/N in (b) reflects the S/N tier structure from multiple exposures \citep[see][]{Kollmeier2026}; the key stellar labels are $\teff$ (c), $\logg$ (d), $\vsini$ (g) and log$L$ (i); the distribution of reddening parameters is shown in (e) and (f); the spectrogeometric distances (from the spectrophotometry of the XP spectra and the parallax) are shown in (h). The isochrone-inferred masses and ages are shown in (j) and (k), with two simple fiducial lines overplotted: the expectation for a uniform age distribution in (j) and the expectation for a Salpeter initial mass function and a uniform age distribution, accounting for the stars' lifetime at $\teff > 10$kK (k).}
\end{figure*}

Fig.~\ref{fig:1Dhistograms} presents the one-dimensional distributions of the sample's properties and fitted parameters. These come in several groups among the sub-panels of Fig.~\ref{fig:1Dhistograms}. The apparent magnitudes and spectral signal-to-noise ratio in (a) and (b), show that the sample spans almost three orders of magnitude in observed flux at S/N$\gtrsim 30$. The fundamental stellar parameters \teff\, and \logg\, are shown in (c) and (d); the limit in \teff\, at 10kK is by construction, and hotter stars are of course less common: of the \Nfiducial\, stars there are $5{,}919$ (9.2\%) with \teff\, above 20kK, $654$ (1.0\%) above 30kK, and 17 above 40kK; the range of \logg\, is limited by the requirement of (\teff,\logg) being near isochrones. The stars' effective $\vsini$ is shown in (g), with a median of 152~km\,s$^{-1}$, and 18\% of the stars with $\vsini > 250$~km\,s$^{-1}$, broadly in line with expectations for a hot star sample; given the modest resolution of the BOSS spectra, the $\vsini$ distribution has an effective floor at $\sim 85$~km\,s$^{-1}$ (see Sect.~\ref{sec:val_external}). The extinction properties are shown in sub-panels (e) and (f): despite being an optically-selected sample, the median extinction is 2.66~mag, with over $5{,}000$ stars at $A_V>5$~mag and $\sim\!2{,}000$ beyond $A_V=6$~mag; 81\% of the stars have extinction-curve slopes $R_V$ within 0.5 of the canonical 3.1, in part reflecting the weak $R_V=3.1\pm0.3$ prior of the fit (Sect.~\ref{sec:params}).  The distribution of spectrogeometric distances, $d$, is shown in panel (h) and illustrations the spatial reach of the sample throughout the Galactic disk: the median distance is 3.4~kpc, with $\sim 56{,}000$ stars $>$2~kpc, $\sim 14{,}000$ stars $>$5~kpc, and still $\sim 2{,}700$ stars $>$8~kpc, unmatched by any other spectroscopic hot-star sample for the Galactic disk. As panel (i) shows, the stars span over three orders of magnitude in bolometric luminosity, with the lower cut-off by sample design (cutting out e.g. hot subdwarfs etc.). 

This brings us to the astrophysically important 
\emph{isochrone-inferred} quantities, the ages $\tauage$ (j) and (initial) masses $M$ (k). Here it is important to remember (Sect.~\ref{sec:fitvsiso}) that the $\tauage$ estimates are posterior estimates with a uniform age prior: so for a star on the MS, its age is effectively half its MS lifetime. The age cut-off in (j) above $\log{\tauage}=8.5$ is of course because such stars are not hot enough. The mass distribution is shown in (k), with a median mass of 4.5\msun , $6{,}544$ stars $>8\msun$ (massive enough to go supernova), $\sim 1{,}400$ stars $>15\msun$, and $\sim 130$ stars $>30\msun$. By temperature, $654$ stars are hotter than $30\kK$, and therefore in the O-star regime; $\sim\!6{,}500$ exceed $8\msun$.

\begin{figure}
\includegraphics[width=\columnwidth]{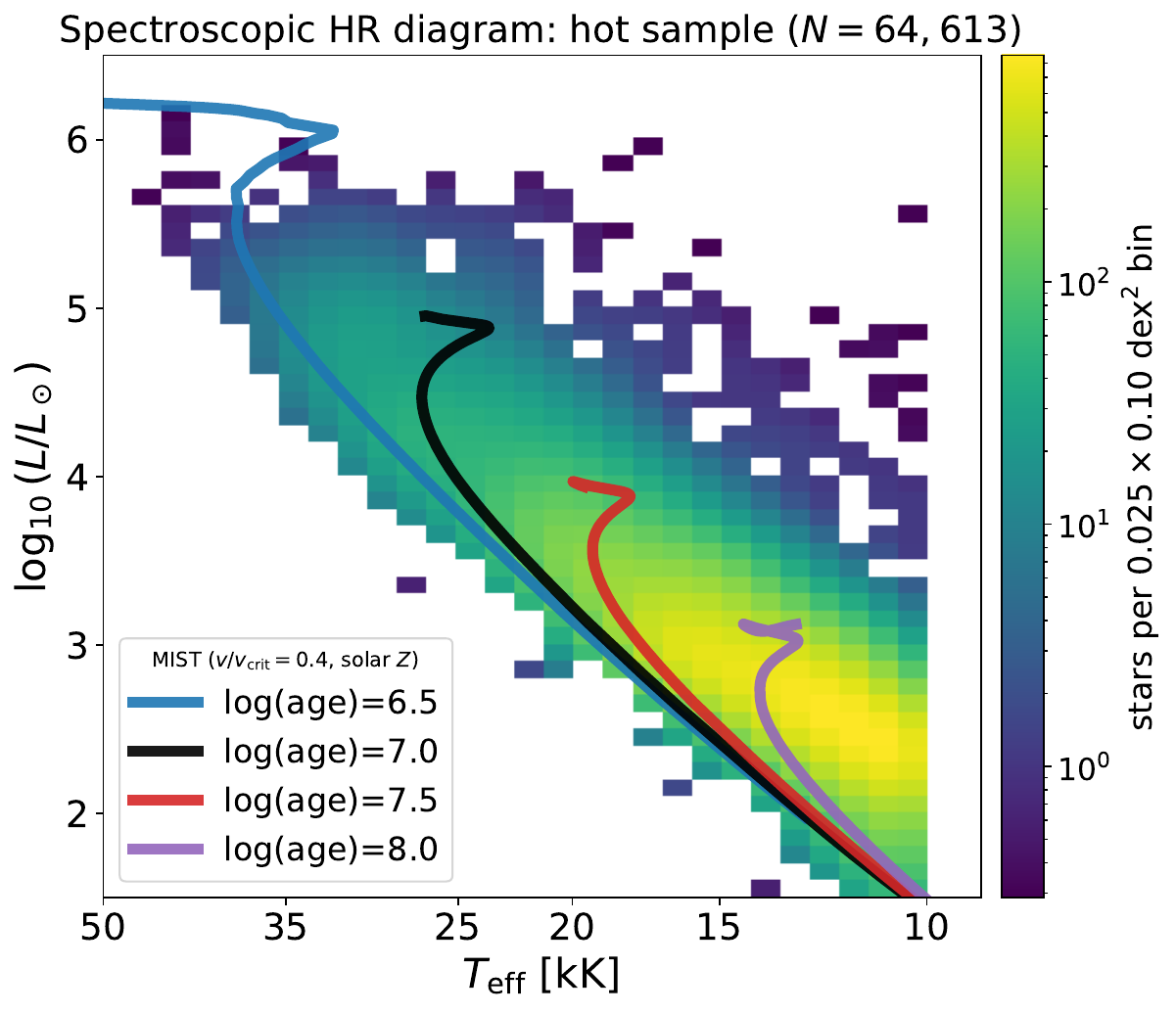}
\caption{\label{fig:hrd}Hertzsprung--Russell diagram of the hot sample ($\teff$ versus $\log L$;
density) with rotating MIST isochrones (solar $Z$, $v/v_{\rm crit}=0.4$, main-sequence portions; $\log{\rm age}=6.5$--$8.0$). The stars trace the expected
young main sequence and turn-off. The map bins the sample in $0.025\times0.10$ dex$^2$ cells (only
cells containing at least one star are coloured) and is convolved with our estimates of realistic per-star
uncertainties ($0.03$ dex in $\log\teff$, $0.13$ dex in $\log L$).}
\end{figure}

\begin{figure}
\includegraphics[width=\columnwidth]{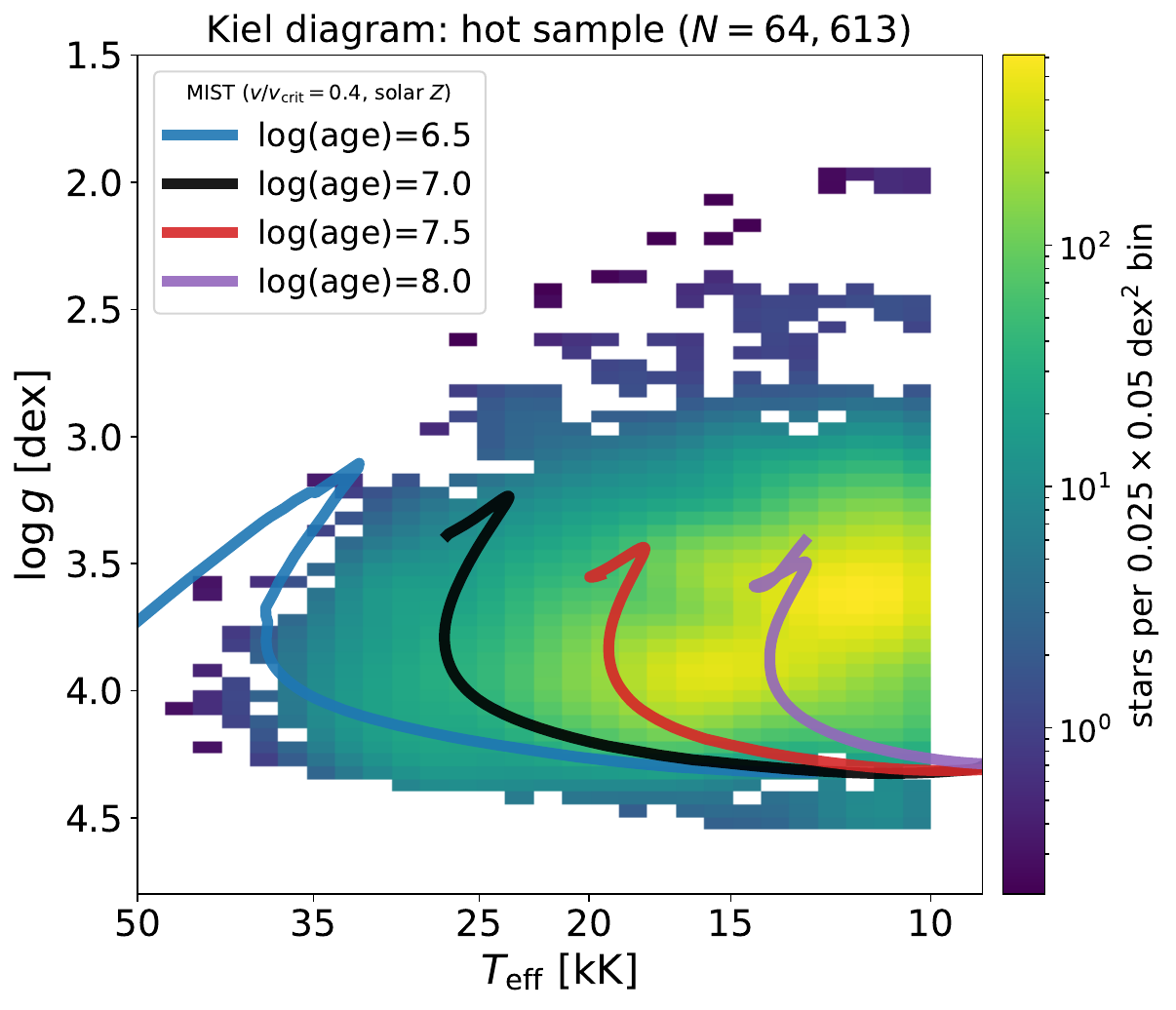}
\caption{\label{fig:kiel}Kiel diagram ($\teff$ versus $\logg$) of the hot sample, rendered exactly as
the HR diagram (Fig.~\ref{fig:hrd}): $0.025\times0.05$ dex$^2$ bins with only $\ge\!1$-star cells
coloured, convolved with our estimate of the realistic uncertainties ($0.03$ dex in $\log\teff$, $0.10$ dex in
$\logg$).}
\end{figure}

In the Hertzsprung--Russell diagram (Fig.~\ref{fig:hrd}), the stars of the fiducial sample  occupy the upper main
sequence and its turn-off and fall in the area covered by solar-metallicity MIST isochrones, as they must given the sample selection cuts and the
generative picture for (\teff , \logg ) of Sect.~\ref{sec:fitvsiso}. The Kiel diagram (Fig.~\ref{fig:kiel}) spans the range from the
$\logg\approx4.3$ main sequence to lower-gravity evolved stars, with the hottest objects at the gravities expected for O stars. Both diagrams also show finer localised over-densities; part of this structure reflects the discrete model grid and isochrone nodes (Sect.~\ref{sec:models}), and we caution that not all of it is yet fully understood.

The width of the main sequence in both diagrams, a \emph{band} rather than a line, is expected rather
than anomalous. A star spends roughly a fixed time per unit of hydrogen burned, and as its core converts
H to He the rising mean molecular weight makes it brighten and its envelope expand; for massive stars the
large (overshooting-enhanced) convective core makes this drift substantial, so the main-sequence band
spans $\sim\!0.1$\,dex in $\log\teff$ and a factor $\sim\!2$--$3$ in luminosity before core-hydrogen
exhaustion. The ZAMS is only the hot edge of this band: for the continuous star formation expected over
the $5$--$30$\,Myr lifetimes shown, stars fill the band roughly uniformly in fractional main-sequence
age, with only the first $\sim\!10$--$15\%$ of the lifetime lying within $\sim\!0.02$\,dex of the ZAMS
(for a solar-mass star the band is so thin that ``main sequence'' and ``ZAMS'' coincide; for $15\,\msun$
they do not). Our magnitude limit reinforces the impression of a sparsely-populated hot edge: at higher
\teff\ the bolometric correction is larger, so a fixed $G$ limit selects higher luminosities and
preferentially removes the lower-mass stars that would sit near the ZAMS at $\log L\approx4$. A residual
deficit of the very youngest stars, the ``ZAMS gap'' discussed by \citet{Castro2014} and
\citet{Holgado2020}, may additionally reflect a brief, optically-obscured early phase or imperfections
in the adopted \teff\ scale and convective-overshooting prescription; that part is not settled, but the
bulk of the band is the straightforward consequence of main-sequence evolution and selection.

We note one systematic at the cool edge of the sample: near the $10$--$11\kK$ floor the median gravity
is relatively low ($\logg\approx3.6$) and the same stars sit somewhat high in the HR diagram. These are
not two effects but one, because at fixed \teff\ the fitted $\logg$ and $\log L$ are tightly anti-correlated
($r\approx-0.9$), both set by the SED. Therfore, the modest $\sim\!1\kK$ warm bias of our temperatures at the
late-B/A boundary (Sect.~\ref{sec:val_bossnet}) inflates the inferred radius and luminosity and lowers
$\logg$ in lockstep. The effect is confined to the coolest ``hot'' stars and does not affect the bulk
of the sample at $\teff\gtrsim13\kK$.

\subsection{Comparison to existing hot-star samples}\label{sec:perspective}
It is worth placing the catalog in the context of existing \emph{spectroscopic} hot-star samples. High-resolution, model-atmosphere
studies such as IACOB \citep{Holgado2020} deliver the most accurate individual parameters but for
a few hundred O-stars, selected for brightness and observability (our own $\sim\!650$ stars above $30\kK$ still exceed that O-star census by a factor of $\sim\!2$); our catalog trades a certain level of
per-star fidelity for an overall sample three orders of magnitude larger, analysed uniformly, and reaching to
the fainter, more reddened, more distant stars that dominate the Galactic OB population. The closest
comparison in size is the LAMOST OB-star work of \citet{Xiang2022}, with which we agree on the basic stellar labels at the
few-percent level (Sect.~\ref{sec:val_external}); compared to LAMOST, SDSS-V's biggest advantage is the uniform coverage across the Galactic plane. What also distinguishes the present catalog from the LAMOST analysis is the
\emph{joint} use of the absolute \emph{Gaia} XP flux and the parallax, which supplies a
spectrophotometric luminosity and hence a mass for every star, independent of the spectroscopic
gravity (Sect.~\ref{sec:fitvsiso}).

\subsection{Products for downstream analyses}\label{sec:products}
Table~\ref{tab:catmain} shows a small sub-sample of the fiducial hot-star catalog with its key
columns. The complete DR20 catalog is available as an electronic table with 77 columns,
including $1\sigma$ uncertainties, with every column defined in Appendix~\ref{sec:appC}.
In line with the data-release scope of this paper (Sect.~\ref{sec:sampledef}), it
contains the $149{,}714$ fitted stars whose co-added spectra are published in SDSS
DR20; the fiducial sample is the subset selected by the flags of
Sect.~\ref{sec:sampledef}. The derived stellar parameters are released together with
this paper; the BOSS epochs underlying each co-add are public in DR20 for the
(majority) pre-cutoff visits, while any later epochs will appear only with SDSS-V Data
Release~22. The full catalog is also archived at Zenodo under
DOI~10.5281/zenodo.23913922.

\begin{deluxetable*}{lrrrrrrrrrrrr}
\tabletypesize{\scriptsize}
\tablecaption{The fiducial hot-star (OB) catalog: key columns (stub). Units: RA, Dec in deg; $G$, $A_V$ in mag; $T_{\rm eff}$ in K; $v\sin i$ in km\,s$^{-1}$; $d$ in pc; $M$ in $M_\odot$; $\log g$, $\log L$, $\log\tau$ in dex.\label{tab:catmain}}
\tablehead{\colhead{\texttt{sdss\_id}} & \colhead{Gaia DR3} & \colhead{RA} & \colhead{Dec} & \colhead{$G$} & \colhead{$T_{\rm eff}$} & \colhead{$\log g$} & \colhead{$A_V$} & \colhead{$v\sin i$} & \colhead{$d$} & \colhead{$\log L$} & \colhead{$M$} & \colhead{$\log\tau$}}
\startdata
103932317 & 5891734023849296512 & 219.4093 & -56.8115 & 13.50 & 10343 & 3.86 & 2.14 & 355 & 2062 & 2.20 & 3.2 & 8.36 \\
54473085 & 428606954121083776 & 7.9500 & 60.5682 & 10.60 & 14458 & 3.18 & 1.69 & 162 & 3042 & 3.56 & 7.0 & 7.65 \\
54518323 & 431227743166936064 & 4.2159 & 62.2290 & 14.80 & 15694 & 3.56 & 3.28 & 84 & 7180 & 3.32 & 6.2 & 7.70 \\
65969285 & 2018254477287951104 & 293.5401 & 21.3310 & 14.75 & 10368 & 3.59 & 3.79 & 236 & 3071 & 2.36 & 3.4 & 8.34 \\
83192403 & 4067023176761885824 & 267.9063 & -25.5926 & 12.18 & 16472 & 3.27 & 4.06 & 97 & 2916 & 3.87 & 8.6 & 7.49 \\
\enddata
\tablecomments{Representative columns for the first five stars of the fiducial sample; $1\sigma$ uncertainties and the remaining columns are in the electronic table (\texttt{sdssv\_galactic\_ob\_catalog\_DR20\_Nov18.2026.fits}/\texttt{.csv}; definitions in Appendix~\ref{sec:appC}).}
\end{deluxetable*}

\begin{figure}
\centering
\includegraphics[width=\linewidth]{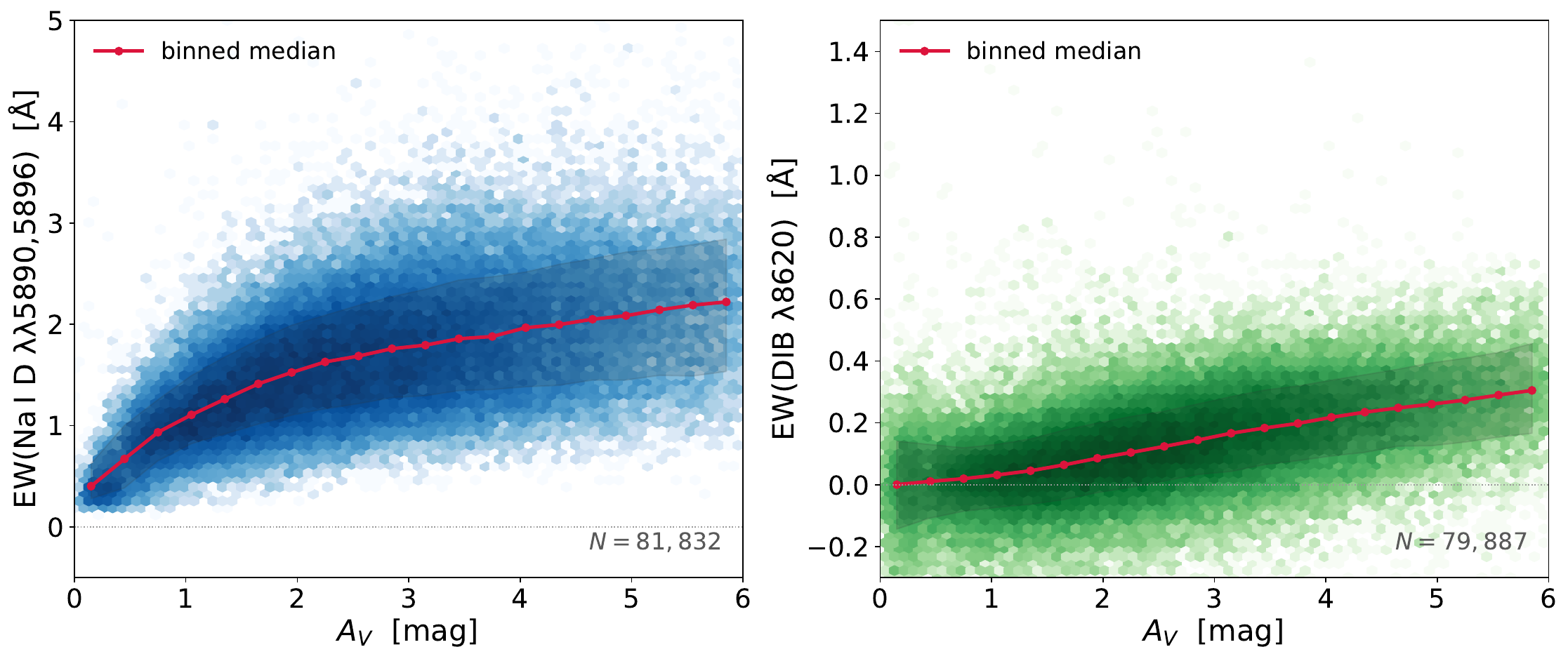}
\caption{Interstellar absorption versus fitted extinction on the hot ($\teff>12\kK$) sightlines, where
the stellar contribution to both features is negligible. \emph{Left:} equivalent width of the Na\,D
$\lambda\lambda$5890,5896 doublet; \emph{right:} equivalent width of the $\lambda$8620 diffuse
interstellar band. The red curve is the binned median with a 16--84\% band over the two-dimensional
density. Na\,D saturates, whereas the $\lambda$8620 band remains near-linear, making it the cleaner tracer of
the reddening column. Both are measured in windows that are masked from the atmospheric fit itself.}
\label{fig:ism}
\end{figure}

Beyond the stellar atmospheric parameters, the catalog also contains spectral measurements and quantities that may be useful in their own right. For each star we tabulate equivalent widths (EWs) of the principal diagnostic lines---the Balmer lines (H$\alpha$, H$\beta$, H$\gamma$), seven He\,{\sc i} and three He\,{\sc ii} lines, Ca\,{\sc ii}\,K, the red Ca\,{\sc ii}\,8542 line and Paschen\,11 ($\lambda$8863)---as measured on the observed spectrum. These EWs are a compact, model-independent summary of the spectrum: they drive the hot/cool gate (Sect.~\ref{sec:sampledef}) and provide an independent check on the atmospheric parameters (Sect.~\ref{sec:val_bossnet}). From the far-red Paschen and Ca\,{\sc ii} EWs we additionally form a reddening-insensitive temperature index, $(\mathrm{Pa}-\mathrm{Ca\,II})/(\mathrm{Pa}+\mathrm{Ca\,II})$, which rises monotonically with \teff\ and retains diagnostic power for the most heavily reddened stars. We also
report two interstellar diagnostics extracted from features that fall within
the fitting windows (but are masked for the stellar label fits): the equivalent widths of the Na\,D doublet and of the $\lambda$8620 diffuse interstellar band (with quality flags). We also measure their velocities, which we use below for internal consistency checks but do not release in the catalog. These quantities are measured on the coadd and
the velocities restored to the heliocentric frame via \texttt{XCSAO\_RV}; because the coadd is built
in the \emph{stellar} rest frame, a star whose radial velocity varied between epochs smears the
stationary interstellar line across the stack, which can bias its measured EW slightly too high. As most of
our stars show little epoch-to-epoch velocity variability relative to these lines' widths, this is a
small effect, but it may affect a subset of binaries. Both trace the intervening medium and
correlate with the fitted extinction, as Fig.~\ref{fig:ism} shows. They provide an independent reddening indicator alongside $A_V$: the Na\,D doublet rises steeply with $A_V$ and then saturates. In contrast,  the unsaturated $\lambda$8620
band grows near-linearly with $A_V$ and is thus the cleaner column tracer of the two. The Na\,D and DIB velocities carry the kinematic imprint of the intervening gas: in the Local-Standard-of-Rest
frame they trace the differential rotation of the intervening medium along each sightline, as shown e.g. in
dedicated DIB kinematic studies \citep[e.g.][]{Zasowski2015,Puspitarini2015,Zhao2021,GaiaDR3DIB2023}.
These two ISM features fall in the blue  ($\lambda$5890) and far-red ($\lambda$8620) ends of the spectrum, so
their agreement is also an internal wavelength check: adopting the literature DIB rest wavelength
($8620.8$\,\AA\ air; \citealt{GaiaDR3DIB2023}) the two interstellar velocities agree in the median to
$0.3\kms$ over the $\sim\!2700$\,\AA\ baseline, confirming that the BOSS wavelength solution is consistent across the full spectrum. Finally, we provide a helium-only
estimate of $\vsini$ as a cross-check on
the full-spectrum value. For the radial velocity our recommended science value is
\texttt{XCSAO\_RV} from the BOSS pipeline. Our joint fit does solve internally for a stellar velocity, but
because the coadds are built in the star's rest frame this fitted velocity is only a residual scattered
about zero (Sect.~\ref{sec:boss}); it carries no stellar-kinematic information, so we do \emph{not} publish it.

To decouple the luminosity from the adopted distance, we also tabulate \texttt{logFbol}, the logarithm
of the dereddened apparent bolometric flux ($\log_{10}F_{\rm bol}$, with $F_{\rm bol}$ in
erg\,s$^{-1}$\,cm$^{-2}$). The luminosity at any adopted distance $d$ then follows from
\begin{equation}
\log_{10}(L/L_\odot) = \texttt{logFbol} + 2\log_{10}(d/{\rm kpc}) + 10.495,
\label{eq:logfbol}
\end{equation}
where the additive constant is $\log_{10}[4\pi(1\,{\rm kpc})^2/L_\odot]$. Evaluated at the parallax
distance this returns the direct photometric luminosity, \texttt{logL\_obs}, which for the fiducial sample lies on average
$0.06$\,dex above the isochrone-posterior \texttt{logL\_fit} (16th--84th percentile range $0.04$--$0.11$\,dex); it is the
natural quantity to use whenever an external distance is preferable to the parallax, as for instance the
known distance to a Magellanic Cloud member. Because \texttt{logFbol} is a measured flux rather than an
isochrone-posterior luminosity, it is also substantially less exposed to model-grid discreteness than
\texttt{logL\_fit}: \texttt{logL\_fit} is effectively a function of the fitted $(\teff,\logg)$, read off
the isochrones, and so maps the discrete isochrone and (through those parameters) spectral grids into
luminosity, whereas \texttt{logFbol} is a continuous observable that sees those
grids only through a slowly-varying bolometric correction (empirically, at fixed $(\teff,\logg)$ the
implied luminosity varies about four times more for \texttt{logFbol} than for \texttt{logL\_fit},
reflecting the genuine star-to-star flux rather than the grid). The complete column list is given in
Appendix~\ref{sec:appC}.

\begin{figure*}
\includegraphics[width=\textwidth]{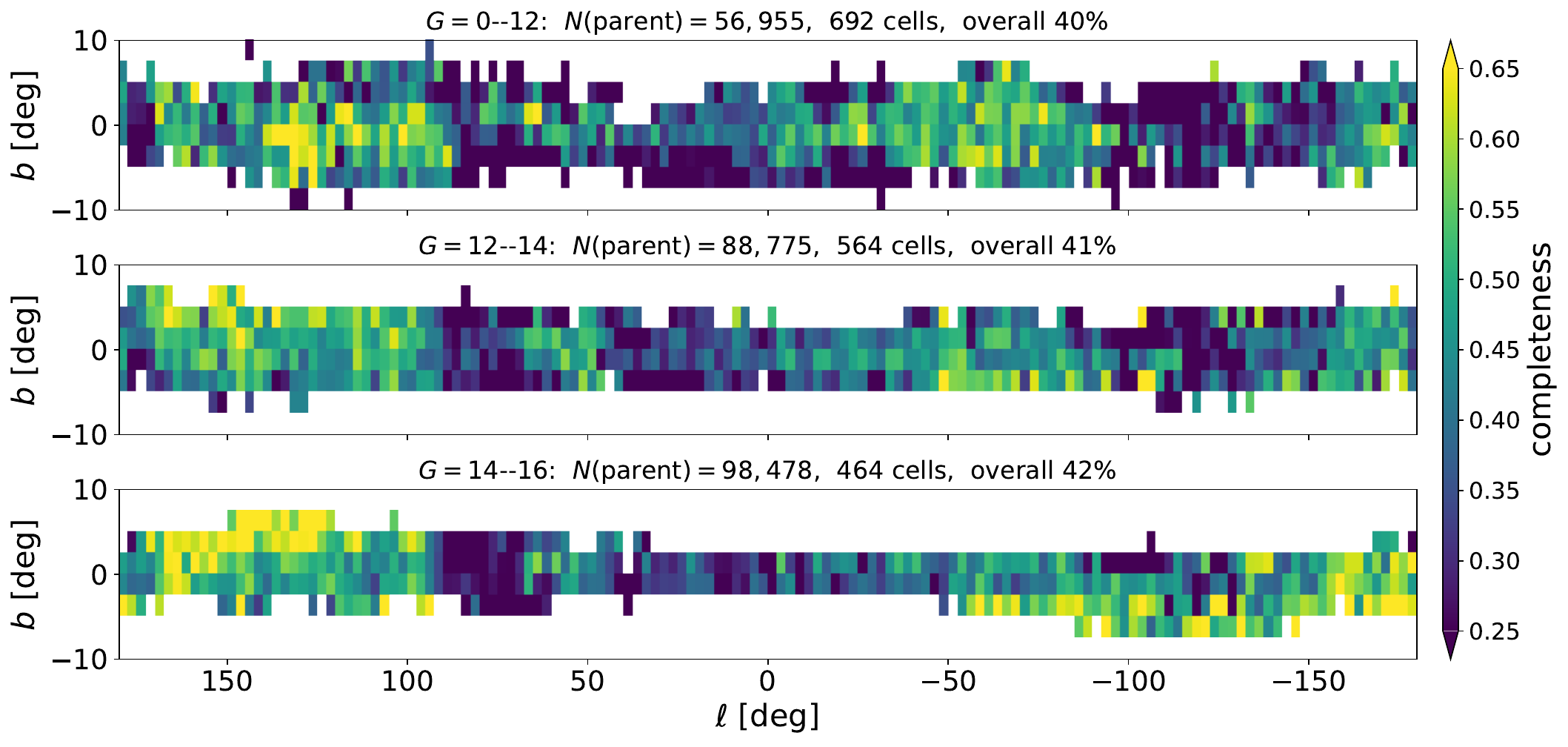}
\caption{\label{fig:complete}Fraction of presumably-hot ($P(\teff>10\kK)>0.5$) parent-catalog stars
that have good BOSS parameters, in equal-area $(\ell,b)$ grid cells at $|b|<10\degr$, split by \emph{Gaia}
$G$ magnitude. Completeness is $\sim40\%$ overall, roughly flat with magnitude, and somnewhat patchy in position.}
\end{figure*}

\section{Selection Function and Completeness}\label{sec:selfunc}
The selection function of any sample describes the probability of any object with given properties (luminosity, temperature, 3D position etc.) to end up in a "final" sample, such as our fiducial sample. Knowing the selection function is indispensable for (forward) modelling the ensemble properties of this sample, based on some physical model; determining the selection function is often a complex multi-step process \citep[e.g.][]{Rix2021}. Broadly speaking, the selection function determination for the case at hand is split into two steps: first, we need to spell out what properties objects would have to have to end up in the initial sample of \citet{Zari2021} that formed the basis of the OB star target carton in SDSS-V; there we have, by design of the catalog, a relatively straightforward answer: for hot stars ($\teff > 10$kK) that are brighter than $G=16$~mag, the initial target catalog is close to complete (just with many contaminating cooler stars).  Second, we need to quantify what the chances were that an object in the target catalog ended up in our fiducial spectroscopic sample. This means that it had to be targeted and observed, and the resulting catalog entries allowed it to pass all the fiducial catalog cuts.  Because a candidate is observed only if the SDSS-V robotic fiber system ends up assigning it a fiber in a scheduled field (Sect.~\ref{sec:targets}), this aspect of the sample's selection function is set operationally rather than by a simple cut in observable quantities (such as $G<16$~mag); hence we must characterize it empirically.

 In defining the selection function for our fiducial hot-star sample, we have two choices that answer different questions: one could normalise this sample by the \emph{full} \citet{Zari2021} target catalog, taking it to be the fraction of all initial candidates that were observed and passed our cuts, as a function of magnitude and position. This measures the raw operational efficiency of the fiber assignment, but it is dominated by the numerous cool ``contaminants" in that catalog. What a user of \emph{this} catalog instead ideally would want is the completeness of the \emph{hot} stars in the initial catalog. In principle the two can differ, as fiber-assignment incompleteness is worst in crowded regions, and the young, hot stars are more spatially clustered than the predominantly older contaminants, which could depress their completeness in dense fields. Empirically, however, we find the net completeness to be essentially flat with on-sky target density across the crowded Galactic plane. It varies by $\lesssim\!2\%$ between the least- and most-crowded quartiles, and if anything rises slightly toward denser fields; consequently, the two denominators yield very similar answers and the choice is not critical. We adopt the purer of the two and characterise the completeness relative to the parent population
of presumably-hot stars, which is \emph{not} the full, and very contaminated \citet{Zari2021} catalog. The completeness we quote is accordingly \emph{conditional} on membership in this photometric-hot parent (defined next), and does not itself capture incompleteness of the photometric proxy (e.g. from extreme reddening or unresolved binaries). We do this by computing for every star in this parent catalog a purely photometric probability of being hot, $P(\teff>10\kK)$, by dereddening its \emph{Gaia} and 2MASS
colours with the \citet{Wang2025_dust_map} three-dimensional dust map and comparing to the loci of hot and cool
stars; this provides a temperature proxy (which was not available at the time of the \citet{Zari2021} work and SDSS-V targetting) that is independent of whether the star was ever observed spectroscopically. We then approximate the selection probability as the fraction of the presumably-hot parent stars
($P(\teff>10\kK)>0.5$) that end up in the fiducial sample. Concretely, within each $(\ell,b)$ cell we estimate the completeness as the ratio
\begin{equation}
C(\ell,b)=\frac{N_{\rm fid}(\ell,b)}{N_{\rm hot\text{-}parent}(\ell,b)},
\label{eq:completeness}
\end{equation}
where $N_{\rm hot\text{-}parent}$ counts the parent stars with $P(\teff>10\kK)>0.5$ and $N_{\rm fid}$ counts those of them that enter the fiducial sample defined in Sect.~\ref{sec:sampledef} (a successful fit that passes the Table~\ref{tab:gate} helium gate, S/N$>20$, spectrophotometric distance $<20$\,kpc, and is not a Magellanic-Cloud member). Fig.~\ref{fig:complete} maps this fraction across the low-latitude sky.
It is $\sim40\%$ overall, roughly independent of
apparent magnitude and
patchy in Galactic longitude, in part tracing the (ongoing) SDSS-V field coverage. This confirms that the fiber assignment is the dominating process in the selection function, not the apparent magnitude. 

\section{Summary}\label{sec:summary}
\begin{figure*}
\gridline{\fig{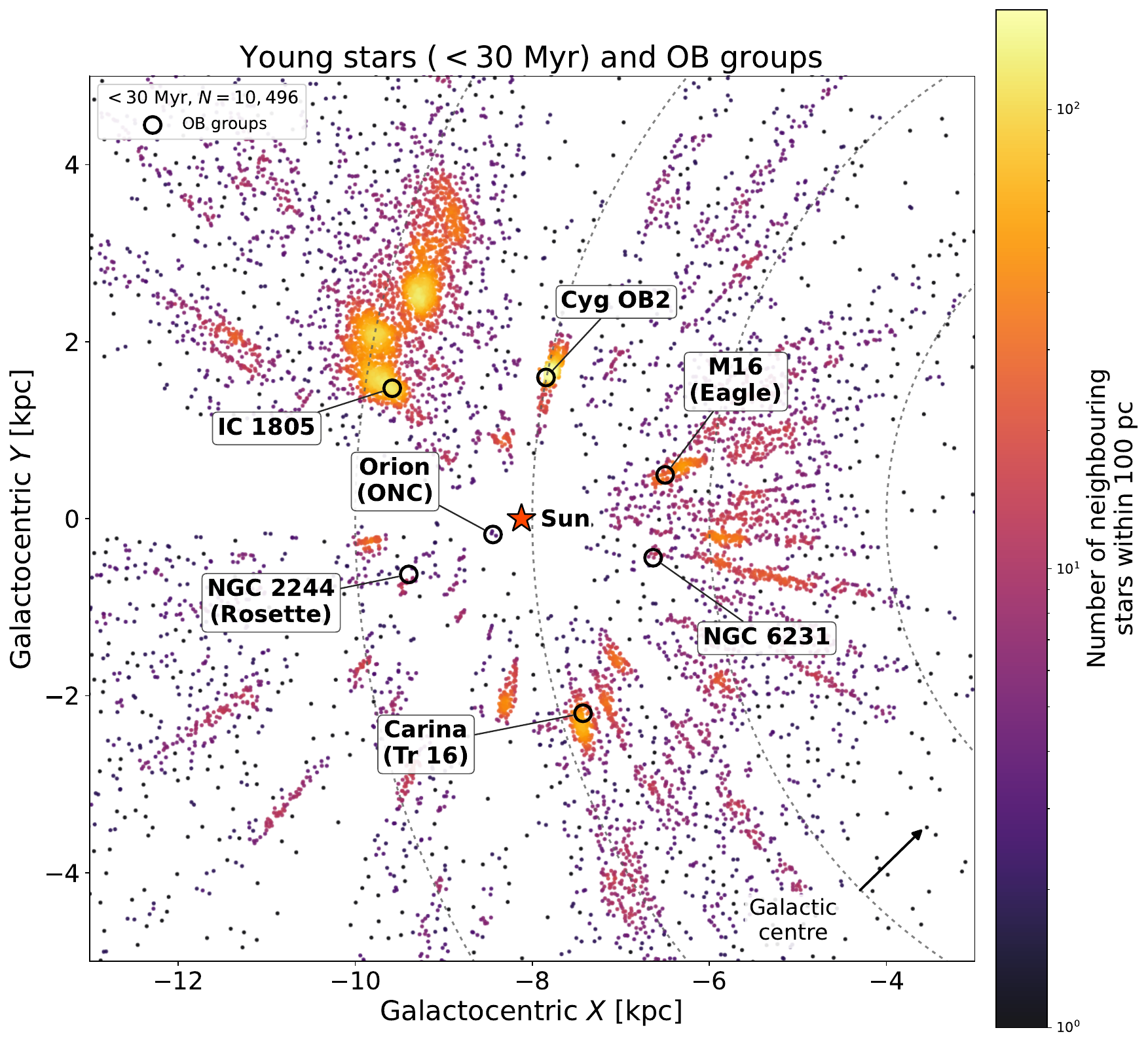}{0.5\textwidth}{(a) young stars}
          \fig{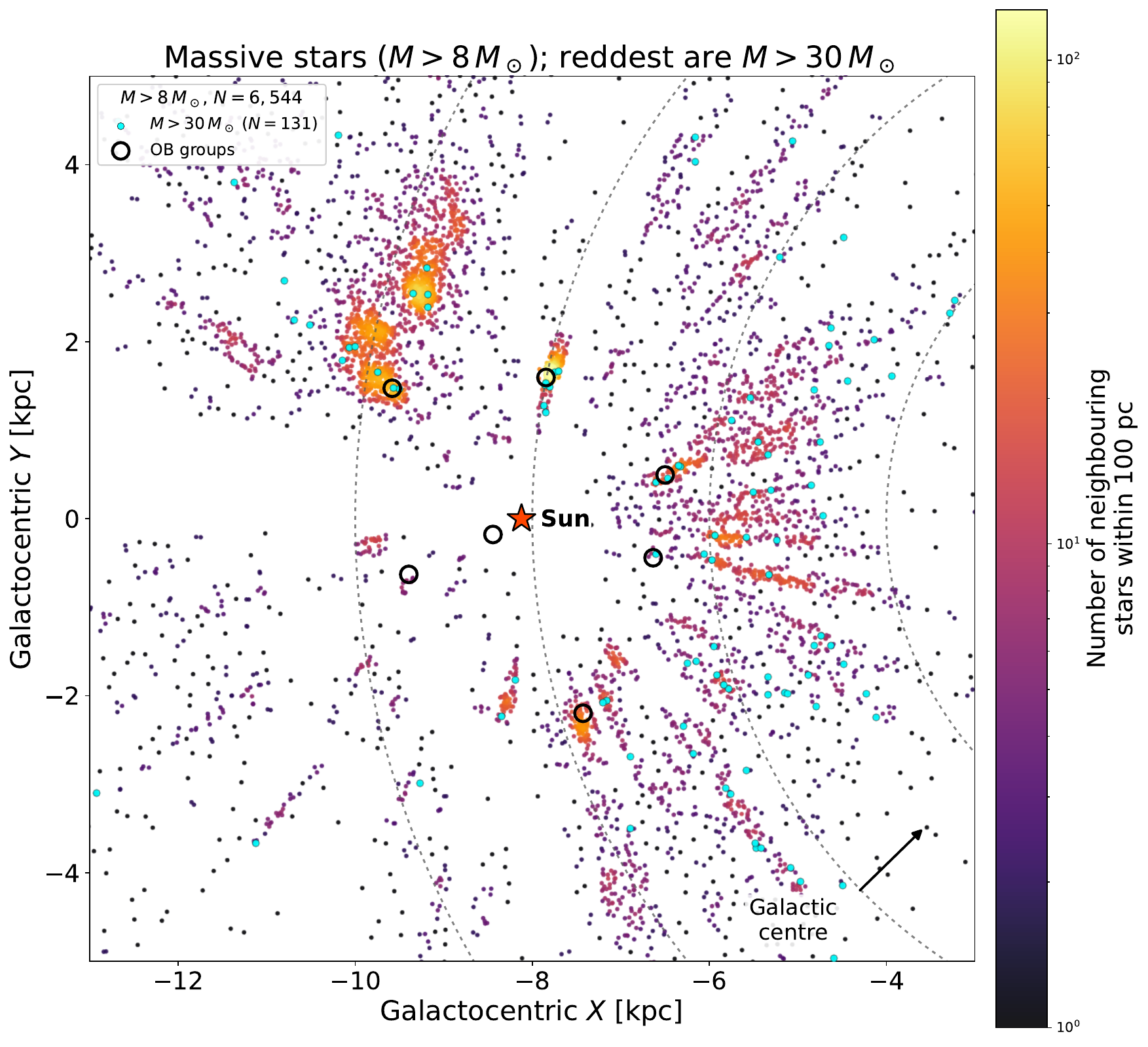}{0.5\textwidth}{(b) massive ($>8\,\msun$) stars}}
\caption{\label{fig:maps}Face-on ($X$--$Y$) maps of the hot sample in Galactocentric coordinates,
illustrating the catalog's reach across the Galactic disk (the Sun is the orange star, the arrow marks
the direction to the Galactic centre, dashed circles are Galactocentric radii). Points are coloured by
local surface density (number of neighbours within $100$\,pc). \emph{(a)} The young ($<30$\,Myr) stars,
which concentrate into the known OB associations and star-forming complexes; the seven most prominent
are labelled (open circles), with Orion and the Rosette as the nearest. \emph{(b)} The massive
($M>8\,\msun$) stars, with the most massive ($M>30\,\msun$) highlighted in cyan. The spiral-arm
segments traced by both populations and the concentration of recent star formation toward the inner
disk are apparent; a quantitative analysis is deferred to companion papers.}
\end{figure*}

We have presented a spectroscopic catalog for $\sim\!\Nhotapprox$ OB stars in the Galactic disk. The value of a catalog of this size and uniformity lies in what astrophysics it enables downstream. We close by
previewing that possible science reach rather than exploiting it here. Because every star has a spectrophotometric
distance, the sample can be placed in three dimensions: Fig.~\ref{fig:maps} shows the young and the
massive hot stars projected onto the Galactic plane within $\sim5$\,kpc, tracing the spiral structure
and the concentration of recent star formation toward the inner disk. These maps are merely illustrative, as we
have applied no correction for the selection function of Sect.~\ref{sec:selfunc}. But they convey
that the catalog resolves the large-scale distribution of the youngest stellar populations across a
substantial fraction of the disk.

Several lines of analysis follow naturally and are the subject of companion papers. The
spectrophotometric distances, combined with \emph{Gaia} proper motions and the XCSAO radial
velocities, yield full six-dimensional phase-space coordinates for a large hot-star sample, enabling a
kinematic census of runaways and of the velocity field of the young disk. The same positions support a
quantitative study of the clustering of recent star formation. The catalog also contains members of
the Magellanic Clouds, fitted with a literature distance prior, which extend the analysis beyond the
Milky Way. In each case the foundational ingredient is the uniform, validated parameter set presented
here.

In summary, we have presented the foundational catalog of the SDSS-V Galactic OB-star program: a uniform
characterisation of the hot stars that SDSS-V has observed spectroscopically, obtained by fitting each
star's \emph{Gaia} XP spectrophotometry, thirteen SDSS-V/BOSS line windows, and the \emph{Gaia}
parallax jointly. The main points are the following.

\begin{enumerate}
\item The joint analysis breaks the temperature--extinction degeneracy that defeats photometry alone
and supplies, for every star, $\teff$, $\logg$, $A_V$, $R_V$, $\vsini$, a radial velocity, and a
spectrophotometric distance, from which masses, ages, luminosities and radii follow through stellar
isochrones. Because the luminosity is \emph{measured} (from the absolute XP flux and the parallax)
rather than inferred from gravity, it is the primary lever on mass wherever the parallax is
informative.
\item Ultimately, we define the hot-star sample by \emph{positive} spectral
identification (helium for hot, metals for cool) rather than by a temperature cut,
yielding $1.2\times10^5$ confirmed-hot stars in two explicit purity tiers ($95{,}443$
with detected helium; $27{,}261$ kept as hot with neither helium nor metals
discernable), of which \Nfiducial\ ($49{,}287$ and $15{,}326$, respectively) are part of the DR20 fiducial sample published here.
\item We add a far-red Paschen $+$ Ca\,{\sc ii}-triplet window that survives interstellar extinction;
for the most reddened stars ($A_V\gtrsim4$), where the blue diagnostics are faint, including this
window reduces the formal $\teff$ and $\logg$ uncertainties by $\sim20\%$ in the median, and its
temperature-sensitive Paschen lines help break the $\teff$--$A_V$ degeneracy.
\item The labels are validated externally against eclipsing-binary dynamical masses
($\approx9\%$), the IACOB and LAMOST temperature scales. And they are validated internally against the SDSS-V BOSSNet
pipeline, which agrees in the overall average but exhibits serious regime-dependent systematics
that a dedicated hot-star pipeline can avoid.
\item We characterise the selection function: the completeness is set by the operational SDSS-V fiber
assignment, is $\sim40\%$ of the presumably-hot parent sample, and is patchy in Galactic longitude.
\end{enumerate}

The astrophysical analysis and interpretation of this catalog is deliberately beyond the scope of this paper: the implications for the IMF, the recent Galactic star-formation-rate, multiplicity and kinematic analyses that this catalog enables will be subject of future work. The main known
limitations of the modelling are: a) the restriction to a single-star analysis; b) the assumption of solar metallicity in both the spectral and isochrone grids (the
radial abundance gradient is shallow, $\sim\!-0.06$\,dex\,kpc$^{-1}$ \citep{Luck2011}. We have accepted this restriction because within the
$\pm3$\,kpc that holds most of our sample the departure from solar stays below the grids' metallicity
spacing; the $>0.5$\,dex spread across the full disk is a worthwhile multi-metallicity check for future
work); c) a small systematic luminosity offset for the
PoWR-fit (hot) stars arises because the PoWR model fluxes are normalised at their native
radius rather than at the isochrone radius used for PHOENIX/BOSZ. This slightly biases \texttt{logL\_obs}
and the isochrone mass for these stars and, d) label uncertainties in the $10$--$11\kK$ regime. All these are documented and flagged.

\vspace{5.8mm plus 1mm minus 1mm}\par\noindent
The authors HWR, JHM and JV acknowledge support from the European Research Council for the ERC Advanced Grant
101054731, ``Hunting Dormant Black Holes in the Galaxy with SDSS-V''.

Funding for the Sloan Digital Sky Survey V has been provided by the Alfred P. Sloan Foundation, the
Heising-Simons Foundation, the National Science Foundation, and the Participating Institutions. SDSS
acknowledges support and resources from the Center for High-Performance Computing at the University of
Utah. SDSS telescopes are located at Apache Point Observatory, funded by the Astrophysical Research
Consortium and operated by New Mexico State University, and at Las Campanas Observatory, operated by the
Carnegie Institution for Science. The SDSS web site is \url{www.sdss.org}.

SDSS is managed by the Astrophysical Research Consortium for the Participating Institutions of the SDSS
Collaboration, including Caltech, The Carnegie Institution for Science, Chilean National Time Allocation
Committee (CNTAC) ratified researchers, The Flatiron Institute, the Gotham Participation Group, Harvard
University, Heidelberg University, The Johns Hopkins University, L'\'Ecole polytechnique f\'ed\'erale de
Lausanne (EPFL), Leibniz-Institut f\"ur Astrophysik Potsdam (AIP), Max-Planck-Institut f\"ur Astronomie
(MPIA Heidelberg), Max-Planck-Institut f\"ur Extraterrestrische Physik (MPE), Nanjing University,
National Astronomical Observatories of China (NAOC), New Mexico State University, The Ohio State
University, Pennsylvania State University, Smithsonian Astrophysical Observatory, Space Telescope Science
Institute (STScI), the Stellar Astrophysics Participation Group, Universidad Nacional Aut\'onoma de
M\'exico, University of Arizona, University of Colorado Boulder, University of Illinois at
Urbana-Champaign, University of Toronto, University of Utah, University of Virginia, Yale University, and
Yunnan University.

This work has made use of data from the European Space Agency (ESA) mission
\emph{Gaia} (\url{https://www.cosmos.esa.int/gaia}), processed by the \emph{Gaia}
Data Processing and Analysis Consortium (DPAC,
\url{https://www.cosmos.esa.int/web/gaia/dpac/consortium}). Funding for the DPAC
has been provided by national institutions, in particular the institutions
participating in the \emph{Gaia} Multilateral Agreement.

Authors CA and AT acknowledge financial support from the Flemish Government under the long-term structural Methusalem funding program by means of the project SOUL: Stellar evOlution in fUll gLory, grant METH/24/012 at KU\,Leuven, from the BELgian federal Science Policy Office (BELSPO) through PRODEX grants for PLATO and {\it Gaia} and from the Research Foundation Flanders (FWO) under grant G089422N.

\par

\software{NumPy \citep{numpy}, SciPy \citep{scipy}, Matplotlib \citep{matplotlib},
Astropy \citep{astropy}, pandas, GaiaXPy.}

\section*{Use of AI tools}
The writing of this manuscript has drawn on the assistance of the Claude LLM 
(Anthropic; Opus and Sonnet models accessed through the Claude Code
command-line assistant). It was used: (i) spelling out first drafts of portions of the manuscript
based on detailed outlines and specifications written by the authors; (ii) writing
and editing extensive portions of the Python/Matplotlib code that created the figures and the summary
statistics reported here, operating solely on the authors' own data, as described; 
and (iii) assisting with literature and reference gathering as a supplement to the NASA ADS.
All resulting text, figures, data, results, or reference was added to this manuscripts only after author review. All scientific content, analysis, and conclusions are the authors' own, and the authors take full responsibility for the content of the figures, data, text, and references in this manuscript. 

\appendix
\section{A-type stars ($7$--$10\kK$)}\label{sec:appA}
Stars whose fit places them at $7000<\teff<10000$\,K fall just below our hot-star floor and are kept
out of the fiducial sample, but their labels remain informative. We tabulate them separately; the full
parameter set is in the electronic catalog (Appendix~\ref{sec:appC}). Table~\ref{tab:appA} lists
representative columns for the first five stars.

\begin{deluxetable*}{lrrrrrrrrrl}
\tabletypesize{\scriptsize}
\tablecaption{A-type stars ($7000<\teff<10000$\,K), 39,443 in total (stub).\label{tab:appA}}
\tablehead{\colhead{\texttt{sdss\_id}} & \colhead{\texttt{gaia\_dr3\_source\_id}} & \colhead{\texttt{ra}} & \colhead{\texttt{dec}} & \colhead{\texttt{phot\_g\_mean\_mag}} & \colhead{\texttt{snr}} & \colhead{\texttt{Teff\_fit}} & \colhead{\texttt{logg\_fit}} & \colhead{\texttt{A\_V\_fit}} & \colhead{\texttt{distance\_pc}} & \colhead{\texttt{model\_source}}}
\startdata
65878314 & 2012557220344611840 & 356.3870 & 60.7571 & 15.74 & 24.3 & 9204 & 3.39 & 4.10 & 4616 & ATLAS \\
100061673 & 5618007339908804480 & 109.6168 & -23.2572 & 14.76 & 29.7 & 8739 & 3.57 & 2.04 & 4237 & ATLAS \\
105803618 & 5972711367273624320 & 258.7645 & -38.6706 & 15.37 & 26.4 & 9347 & 3.50 & 4.32 & 3327 & ATLAS \\
65915769 & 2015008512793417600 & 350.3025 & 62.2576 & 14.13 & 28.1 & 8782 & 3.07 & 3.77 & 3573 & ATLAS \\
81984083 & 4053310553157702784 & 264.0273 & -35.2337 & 15.22 & 23.5 & 9773 & 3.56 & 4.52 & 2729 & ATLAS \\
\enddata
\tablecomments{Representative columns for the first five stars of the A-type ($7000<\teff<10000$\,K) sample; $1\sigma$ uncertainties and the remaining columns are in the electronic table (\texttt{sdssv\_galactic\_ob\_catalog\_DR20\_Nov18.2026.fits}/\texttt{.csv}; definitions in Appendix~\ref{sec:appC}).}
\end{deluxetable*}

\section{Cooler stars ($<7\kK$): identifiers}\label{sec:appB}
Targets that the gate relabels as cool ($\teff<7000$\,K) contaminants are retained for completeness,
but their single-hot-star parameters are not reliable. We therefore provide identifiers only in print;
the full entries, with all flags, are in the electronic catalog. Table~\ref{tab:appB} lists the first
five.

\begin{deluxetable*}{lrrrrr}
\tabletypesize{\scriptsize}
\tablecaption{Cooler stars ($\teff<7000$\,K): identifiers, 31,955 in total (stub).\label{tab:appB}}
\tablehead{\colhead{\texttt{sdss\_id}} & \colhead{\texttt{gaia\_dr3\_source\_id}} & \colhead{\texttt{ra}} & \colhead{\texttt{dec}} & \colhead{\texttt{phot\_g\_mean\_mag}} & \colhead{\texttt{Teff\_fit}}}
\startdata
73407519 & 3003672532215649152 & 97.8916 & -8.5063 & 15.80 & 6855 \\
104756656 & 5935350344433969792 & 245.9681 & -49.9503 & 15.88 & 5522 \\
75858497 & 3219249271647929984 & 85.0149 & -0.3351 & 15.18 & 6399 \\
63761469 & 1822720215631809920 & 302.4379 & 19.9144 & 15.62 & 5781 \\
84246833 & 4096905605041673984 & 276.6674 & -17.1282 & 15.75 & 6627 \\
\enddata
\tablecomments{Identifiers only. The fitted parameters of these cool ($\teff<7000$\,K) contaminants are not reliable; they are provided, with all flags, solely in the electronic table (\texttt{sdssv\_galactic\_ob\_catalog\_DR20\_Nov18.2026.fits}/\texttt{.csv}; Appendix~\ref{sec:appC}).}
\end{deluxetable*}

\section{Catalog column definitions}\label{sec:appC}
Table~\ref{tab:coldef} defines the public catalog columns. The complete table, a machine-readable
column dictionary, and the FITS and CSV files are available from the data archive
(Sect.~\ref{sec:products}).

\startlongtable
\begin{deluxetable*}{l l l}
\tabletypesize{\scriptsize}
\tablecaption{Public catalog columns (77 total). The complete table, a machine-readable
column dictionary, and the FITS/CSV files are available from the data archive
(Sect.~\ref{sec:products}).\label{tab:coldef}}
\tablehead{\colhead{Column} & \colhead{Unit} & \colhead{Description}}
\startdata
\texttt{sdss\_id} & \nodata & SDSS-V identifier; primary join key \\
\texttt{gaia\_dr3\_source\_id} & \nodata & Gaia DR3 source identifier \\
\texttt{ra} & deg & Right ascension (ICRS) \\
\texttt{dec} & deg & Declination (ICRS) \\
\texttt{phot\_g\_mean\_mag} & mag & Gaia DR3 G-band mean magnitude \\
\texttt{parallax} & mas & Gaia DR3 parallax \\
\texttt{parallax\_corrected} & mas & Parallax, zero-point corrected (used in the fit) \\
\texttt{parallax\_error} & mas & Gaia DR3 parallax uncertainty \\
\texttt{pmra} & mas\,yr$^{-1}$ & Gaia DR3 proper motion in RA (pmra*) \\
\texttt{pmdec} & mas\,yr$^{-1}$ & Gaia DR3 proper motion in Dec \\
\texttt{snr} & \nodata & Median S/N of the co-added BOSS spectrum \\
\texttt{has\_boss} & \nodata & Flag: usable BOSS spectrum (else XP-only fit) \\
\texttt{n\_masked\_pix} & \nodata & Number of deviant BOSS pixels masked before the fit \\
\texttt{n\_epochs} & \nodata & Number of BOSS epochs entering the co-add \\
\texttt{n\_epochs\_dr20} & \nodata & Epochs preceding the DR20 cutoff (MJD 60708) \\
\texttt{mjd\_first} & d & MJD of the first epoch \\
\texttt{mjd\_last} & d & MJD of the last epoch \\
\texttt{majority\_before} & \nodata & Flag: at least half of the epochs precede the DR20 cutoff (True for all released rows) \\
\texttt{Teff\_fit} & K & Effective temperature \\
\texttt{Teff\_err} & K & 1-sigma uncertainty on Teff\_fit \\
\texttt{logg\_fit} & dex & Surface gravity log g (cgs) \\
\texttt{logg\_err} & dex & 1-sigma uncertainty on logg\_fit \\
\texttt{A\_V\_fit} & mag & V-band extinction (Gordon et al. 2023 law; allowed range 0-12) \\
\texttt{A\_V\_err} & mag & 1-sigma uncertainty on A\_V\_fit \\
\texttt{R\_V\_fit} & \nodata & Extinction-curve slope R\_V (prior 3.1 $\pm$ 0.3) \\
\texttt{R\_V\_err} & \nodata & 1-sigma uncertainty on R\_V\_fit; NaN when unconstrained \\
\texttt{vsini\_fit} & km\,s$^{-1}$ & Effective $v\,\sin i$: total line broadening, rotation-dominated above about 100 km/s \\
\texttt{vsini\_err} & km\,s$^{-1}$ & 1-sigma uncertainty on vsini\_fit \\
\texttt{vsini\_he} & km\,s$^{-1}$ & Effective $v\,\sin i$ from the helium windows only (preferred for hot stars) \\
\texttt{vsini\_he\_err} & km\,s$^{-1}$ & 1-sigma uncertainty on vsini\_he \\
\texttt{vsini\_he\_flag} & \nodata & Quality flag: 0 reliable, 1 boundary-limited, 2 no usable helium window \\
\texttt{model\_source} & \nodata & Atmosphere family of the best fit: PHOENIX, ATLAS (= BOSZ), or PoWR \\
\texttt{distance\_pc} & pc & Spectrophotometric distance (parallax-constrained) \\
\texttt{logFbol} & log(erg\,s$^{-1}$\,cm$^{-2}$) & Dereddened apparent bolometric flux; log10(L/Lsun) = logFbol + 2 log10(d/kpc) + 10.495 \\
\texttt{logL\_obs} & log($L_\odot$) & Measured luminosity, from the bolometric flux and the parallax distance; NaN if the parallax is uninformative \\
\texttt{logL\_obs\_err} & dex & 1-sigma uncertainty on logL\_obs \\
\texttt{logL\_fit} & log($L_\odot$) & Isochrone-posterior luminosity (consistent with Mass\_fit, logAge\_fit, R\_Rsun\_fit) \\
\texttt{Mass\_fit} & $M_\odot$ & Isochrone-posterior initial mass \\
\texttt{Mass\_err} & $M_\odot$ & 1-sigma uncertainty on Mass\_fit \\
\texttt{M\_spec\_fit} & $M_\odot$ & Spectroscopic mass from log g and R\_Rsun\_fit (independent of the IMF prior) \\
\texttt{logAge\_fit} & log(yr) & Isochrone-posterior age \\
\texttt{logAge\_err} & dex & 1-sigma uncertainty on logAge\_fit \\
\texttt{R\_Rsun\_fit} & $R_\odot$ & Isochrone-posterior radius \\
\texttt{mass\_lever} & \nodata & Observable that constrained the mass posterior: logL (parallax informative) or logg \\
\texttt{n\_iso\_points} & \nodata & Number of isochrone points in the posterior sum \\
\texttt{chi2\_total} & \nodata & Total $\chi^2$ of the joint fit \\
\texttt{chi2\_xp} & \nodata & $\chi^2$ of the Gaia XP spectrophotometry \\
\texttt{chi2\_boss} & \nodata & $\chi^2$ of the BOSS spectral windows; zero for XP-only fits \\
\texttt{chi2\_plx} & \nodata & $\chi^2$ of the parallax term \\
\texttt{cool\_gate} & \nodata & Spectral gate class: confirmed\_hot, hot\_no\_metals, cool, cool\_contaminant, reddened\_ungated, no\_boss \\
\texttt{hot\_confirmed} & \nodata & Flag: passes the hot-star gate (helium, or neither helium nor metals) \\
\texttt{is\_cool\_contaminant} & \nodata & Flag: metal lines present but no helium; relabelled as a cool contaminant \\
\texttt{boss\_bad} & \nodata & Flag: anomalously large BOSS chi-square after masking (corrupt or composite spectrum) \\
\texttt{in\_MCs} & \nodata & Flag: likely Magellanic Cloud member; excluded from the fiducial sample (fit assumes Galactic distance and metallicity) \\
\texttt{mc\_cloud} & \nodata & Magellanic Cloud assignment: LMC, SMC, or empty \\
\texttt{ew\_halpha\_A} & \AA & EW of H alpha 6563 \\
\texttt{ew\_hbeta\_A} & \AA & EW of H beta 4861 \\
\texttt{ew\_hgamma\_A} & \AA & EW of H gamma 4340 \\
\texttt{ew\_heii4686\_A} & \AA & EW of He II 4686 \\
\texttt{ew\_heii4542\_A} & \AA & EW of He II 4542 \\
\texttt{ew\_heii5411\_A} & \AA & EW of He II 5411 \\
\texttt{ew\_hei4026\_A} & \AA & EW of He I 4026 \\
\texttt{ew\_hei4144\_A} & \AA & EW of He I 4144 \\
\texttt{ew\_hei4388\_A} & \AA & EW of He I 4388 \\
\texttt{ew\_hei4471\_A} & \AA & EW of He I 4471 \\
\texttt{ew\_hei4922\_A} & \AA & EW of He I 4922 \\
\texttt{ew\_hei6678\_A} & \AA & EW of He I 6678 \\
\texttt{ew\_hei7065\_A} & \AA & EW of He I 7065 \\
\texttt{ew\_caIIk\_A} & \AA & EW of Ca II K 3933 \\
\texttt{ew\_caII8542\_A} & \AA & EW of Ca II 8542 \\
\texttt{ew\_paschen11\_A} & \AA & EW of Paschen 11 8862 \\
\texttt{ew\_nad\_A} & \AA & EW of interstellar Na D \\
\texttt{ew\_nad\_err} & \AA & 1-sigma uncertainty on ew\_nad\_A \\
\texttt{nad\_flag} & \nodata & Na D quality flag: 0 detected, 1 weak, 2 unavailable \\
\texttt{ew\_dib\_A} & \AA & EW of the 8620\,\AA{} diffuse interstellar band \\
\texttt{ew\_dib\_err} & \AA & 1-sigma uncertainty on ew\_dib\_A \\
\texttt{dib\_flag} & \nodata & DIB quality flag: 0 detected, 1 weak, 2 unavailable \\
\enddata
\end{deluxetable*}

\bibliographystyle{aasjournal}
\bibliography{ms}

\end{document}